\documentclass{SciPost}

\hypersetup{
    colorlinks,
    linkcolor={red!50!black},
    citecolor={blue!50!black},
    urlcolor={blue!80!black}
}

\usepackage[bitstream-charter]{mathdesign}
\DeclareSymbolFont{usualmathcal}{OMS}{cmsy}{m}{n}
\DeclareSymbolFontAlphabet{\mathcal}{usualmathcal}

\fancypagestyle{SPstyle}{
\fancyhf{}
\lhead{\colorbox{scipostblue}{\bf \color{white} ~SciPost Physics Lecture Notes }}
\rhead{{\bf \color{scipostdeepblue} ~Submission }}

\fancyfoot[C]{\textbf{\thepage}}
}

\usepackage{graphicx}
\usepackage{ifthen}
\graphicspath{{pictures/}{programs/}}

\newcounter{privacy}
\newboolean{showall} 
\setboolean{showall}{false}

\newboolean{showcode} 
\setboolean{showcode}{false}

\newcommand{\myvec}[1]{\underline{#1}}

\newenvironment{example}[1]
{\rule{0.999\textwidth}{0.5mm}\vspace{-0.1cm}\begin{quote}
\underline{Example}: #1\\[0.1cm]}
{\hfill$\Box$\\
\end{quote}\vspace*{-0.7cm}\rule{\textwidth}{0.5mm}\\[0.0cm]}

\newenvironment{algorithm}[1]
{
\begin{tabbing} xx \= xx \= xx \= xx \= xx \= xx \= xx \= xxxxx \kill
 {\bf algorithm} #1\\
 {\bf begin}\\
}
{
 {\bf end}\\
 \end{tabbing}
}

\newcommand{\activator}[3]{
 \centerline{\rule{0.42\textwidth}{1pt} [Activator] 
  \rule{0.42\textwidth}{1pt}}
#1 \\
\ifthenelse{\value{privacy}>0}{#3}{#2}
\newline \centerline{\rule{1.00\textwidth}{1pt}} 
}

\newlength{\remarkwidth}
\newcommand{\remark}[2][0.5\textwidth]{
\settowidth{\remarkwidth}{\mbox{\begin{tabular}{l} #21 \end{tabular}}}
\ifthenelse{\lengthtest{\remarkwidth < #1}}%
                       {}{\setlength{\remarkwidth}{#1}}%
\hfill 
\fbox{\begin{minipage}[t]{\remarkwidth}
#2
\end{minipage}
}}
\ifthenelse{\value{privacy}<2}{\renewcommand{\remark}[2][0.5\textwidth]{}}{}

\begin{document}
\pagestyle{SPstyle}

\begin{center}{\Large \textbf{\color{scipostdeepblue}{
        Numerical Aspects of Large Deviations}}}
\end{center}

\begin{center}\textbf{ 
    Alexander K. Hartmann\textsuperscript{1$\star$}}
\end{center}

\begin{center}
  {\bf 1} Institute of Physics, University of Oldenburg, Germany
\\[\baselineskip]
$\star$ \href{mailto:a,hartmann@uni-oldenburg.de}{\small a.hartmann@uni-oldenburg.de}

\end{center}

\section*{\color{scipostdeepblue}{Abstract}}
\textbf{\boldmath{
An introduction to numerical large-deviation
  sampling is provided. First, direct biasing with a known distribution
  is explained. As simple example, the Bernoulli process is used
  throughout the text. 
  Next, Markov chain Monte Carlo (MCMC)
  simulations are introduced. In particular, the Metropolis-Hastings
  algorithm is explained. As first implementation of MCMC,
  sampling of the plain Bernoulli model is shown.
  Next, an exponential bias is used for the same model,
  which allows one to obtain the tails of the distribution of a
  measurable quantity. This approach
  is generalized to  MCMC simulations, where the states
  are vectors of $U(0,1)$ random
  entries. This allows one to use the exponential or any other bias to
  access the large-deviation properties of rather arbitrary random processes.
  Finally, some recent research applications to study more complex models
  are discussed.
}}

\noindent\textcolor{white!90!black}{%
\fbox{\parbox{0.975\linewidth}{%
\textcolor{white!40!black}{\begin{tabular}{lr}%
  \begin{minipage}{0.6\textwidth}%
    {\small Copyright attribution to authors. \newline
    This work is a submission to SciPost Physics Lecture Notes. \newline
    License information to appear upon publication. \newline
    Publication information to appear upon publication.}
  \end{minipage} & \begin{minipage}{0.4\textwidth}
    {\small Received Date \newline Accepted Date \newline Published Date}%
  \end{minipage}
\end{tabular}}
}}
}

\vspace{10pt}
\noindent\rule{\textwidth}{1pt}
\tableofcontents
\noindent\rule{\textwidth}{1pt}
\vspace{10pt}

\section{Motivation}

The aim of theoretical physics is to understand nature by providing models,
which are as simple as possible to explain the targeted phenomena.
Unfortunately, most models cannot be solved analytically. Thus, one has to use
numerical simulations \cite{practical_guide2015} to investigate them.
Here, models for
stochastic processes are considered, which means that quantities
of interest are random variables and characterized by probability
distributions. In this work,
basic methods are introduced, which allow
one to obtain the probability distribution of the quantity of interest
over a large range of the support, down to very small probabilities.

As toy example for the following content,  we use a very simple model,
i.e., the Bernoulli process. This is a series of $n$ independent
coin flips with outcomes $y_i\in \{0,1\}$,
$\myvec{y}=(y_1,\ldots, y_n)$.
Let $\alpha$ be the probability of
obtaining the result $1$ in a single coin flip, i.e.

\begin{equation}
P(y_i=1)=\alpha;\quad P(y_i=0)=1-\alpha\;.
\label{eq:coin:toss}
\end{equation}

Thus, the probability for a certain outcome $\myvec{y}$ is
\begin{equation}
  P(\myvec{y})=\prod_{i=1}^n\alpha^{y_i}(1-\alpha)^{1-y_i}=
  \alpha^{l(\myvec{y})} (1-\alpha)^{n-l(\myvec{y})}
\label{eq:bernoulli}
\end{equation}
where $l=l(\myvec{y})=\sum_i y_i$ is the number of 1's in the $n$-coin
experiment.
Thus, $l$ is independent of the
actual order of 0's and 1's in $\myvec{y}$.
Here, we will be mainly interested in the distribution of $l$.
It is well known that $l$  is distributed according to the
Binomial distribution. Thus, the probability to find for $l(\myvec{y})$
a specific value $l\in \{0,\ldots,n\}$ is given by 

\begin{equation}
P_n(l)= {n \choose l } \alpha^{l} (1-\alpha)^{n-l}\,.
\label{eq:binomial}
\end{equation}
This allows for a simple comparison of the numerical results with the
known distribution.

The present text tries to be very comprehensive, including also very fundamental
material, which many readers might know partially. Thus, a short overview is
given now, to allow the reader maybe to skip some sections. In section two, \emph{simple
  sampling} is introduced, which means that a probability distribution is
sampled according its original underlying distribution. In the best case,
this is done by \emph{direct sampling}, where each call in the program
to a corresponding function
yields an independent realization of the stochastic process.
This is applied to the
Bernoulli process. Then \emph{biased sampling} is introduced, which means that
the system is sampled according to a distribution different from the
original one.
First, the \emph{educated} variant is discussed, where one knows how one has
to ``push'' a system by changing system parameters to sample it in the region of interest.
Again the Bernoulli model is considered in test simulations,
which shows than indeed the tails of the distributions can be reached.

In the third section, \emph{Markov chain Monte Carlo} simulations
are presented. First, Markov chains are introduced and the fundamental
\emph{Master equation} is
explained. Next, a simple Markov chain for the Bernoulli process is built.
Then, the Metropolis-Hasting algorithm is explained, which allows for sampling according to
rather general desired distributions, which is in particular useful
for cases where direct
sampling does not work. Here again, the Bernoulli process is used as example,
for pedagogical reasons, considering that direct sampling is possible there.

In the fourth section, the two main concepts, biasing and Markov chains, are
combined. This allows for ``blind'' sampling, where the process under
investigation is driven into the rare-event region of interest, without the
need to know how exactly the system has to be ``pushed''. This is done automatically
by using a bias where the bias depends on the quantity of interest. Here, the
standard case of an exponential bias, also called exponential tilt, is described.
To demonstrate this, the Bernoulli process is considered. Here, unusual
quantities of measurements, like the number of consecutive equal coin flips of
length of at least 3 are used. This is useful, because the educated sampling
does not work for such quantities.
Then it is explained, how to obtain the true distribution,
i.e., how to remove the bias from the sampled data. This involves also the
determination of normalization constants, i.e., partition functions. In the last
part, it is explained how the basic
algorithm presented so far for the Bernoulli process can be generalized.
This yields a ``black box'' approach, which means
that it is applicable for rather arbitrary stochastic processes, at least if
the quantity of interest is a scalar variable.
Finally, some examples of application
of the large-deviation approach to a variety of models are shortly discussed.

\section{Simple and Biased Sampling}

The most natural implementation of a stochastic system is to
sample the states or configuration   according to the
original distribution of the system. Performing measurements
means to take just the configurations, calculate the measurable quantities,
and then obtain averages or histograms. This is introduced in the first
subsection of this chapter. Next, as an example the Bernoulli process
is considered. Then it is shown that one can reach the rare events, i.e.,
the tails of the distributions, by using biased sampling. Finally,
this will be illustrated for the Bernoulli process again.

\subsection{Simple Sampling \label{sec:simple:sampling}}

In principle, we consider any model which is described by
probabilities $P(\myvec{y})$ of ``objects'' $\myvec{y}$,
e.g, outcome of coin flips,
other random numbers, orientation of spins, position of particles, etc.

We want to measure numerically sample averages of
measurable quantities $A=A(\myvec{y})$. The sample averages are estimates
of expectation values
\begin{equation}
\langle A \rangle := \sum_{\myvec{y}} A(\myvec{y}) P(\myvec{y})\,.
\label{eq:ph-average}
\end{equation}

As numerical approach, we could use \emph{simple sampling},
i.e., generate $M$ configurations 
$\myvec{y}^{(1)},$ $\ldots,$ $\myvec{y}^{(M)}$ 
according to the original probabilities $P(.)$.
For an experiment, this corresponds to just taking measurements
for the given system, as it is.
In this case, we have to calculate a sample to obtain an
estimate for the expectation value
$$
\langle A \rangle \approx \frac 1M \sum_i A(\myvec{y}^{(i)})\,.
$$
In some cases, one can perform the simple sampling in a \emph{direct} way.
This means that one can implement a function or subroutine, which each time
it is called, it generates an independent sample of a configuration,
which is properly distributed. One example is the \emph{inversion method},
which is based on applying \cite{practical_guide2015} the inverted
cumulative distribution
function to a random
number $r$, which is uniformly distributed in the interval $[0,1]$.
For the latter one, one writes $r\sim U(0,1)$.
If on the other hand
no direct sampling approach is available for the considered model,
one has to use more involved techniques, in particular Markov chain
Monte Carlo simulations, see Chap.~\ref{sec:MC}.

\subsection{Simple Sampling for the Bernoulli Process}
\label{sec:simple:Bernoulli}

Let's perform the simple sampling for the Bernoulli process
with parameters $\alpha$ and $n$,
as defined in Eq.~(\ref{eq:bernoulli}). From the
algorithmic point of view, we can use direct sampling.
This means, we iterate over all entries $y_i$
of $\myvec{y}$ and each time we draw a $U(0,1)$ uniformly distributed
number $r$ and if $r<\alpha$ we assign $y_i=1$ and if $r\ge \alpha$
we assign $y_i=0$. Then 
we measure the number $l=l(\myvec{y})=\sum_{i=1}^n y_i$
of 1's. We repeat this many times and build a histogram for the
number of observed occurrences of the different values of $l$.
Here we expect to obtain approximately
the Binomial distribution Eq.~(\ref{eq:binomial}).

To comply with Eq.~(\ref{eq:ph-average}), the histogram is represented 
for $k=0,1,\ldots,n$ by the measurable  quantities
\begin{equation}
  A_l(\myvec{y})=\delta_{l,l(\myvec{y})}=\delta_{l,\sum_i y_i}\,.
  \label{eq:histogram_entry}
\end{equation}
Thus, the sample mean of $A_l$, also called the \emph{empirical measure},
approximates the
probability $P_n(l)$ to measure a number $l(\myvec{y})=l$
of 1's in an experiment with $n$ coin flips.

A C code implmentation is provded in the source code \cite{bernoulli_code2024}
\verb!bernoulli_direct.c!.\footnote{It can be compiled in a unix
shell with {\tt cc -o bernoulli bernoulli\_direct.c -lm}.}

\begin{figure}[ht]
\includegraphics[width=0.6\textwidth]{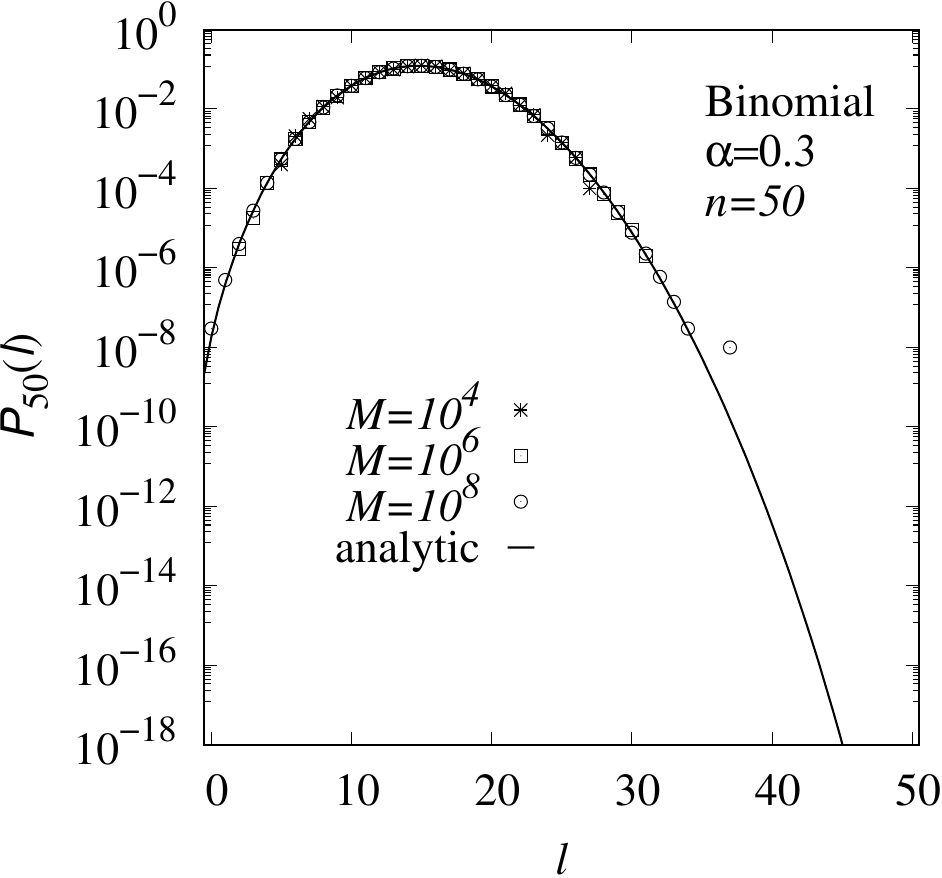}
\caption{\label{fig:binomial:simple}
Numerically measured  histograms of the number $l$ of 1's obtained from  $n=50$
coin flips. The histograms are measured
from $M=10^4$, $10^6$ and $10^8$ experiments.
The solid line shows the analytical solution Eq.~(\ref{eq:binomial}).}
\end{figure}

The result obtained within a simple simulation,
here for $n=50$ and $\alpha=0.3$, could look like shown
in Fig.~\ref{fig:binomial:simple}. We consider three different values
for the number $M$ of samples, namely
$M=10^4$, $10^6$ and $10^8$.
The histograms follow the true distribution Eq.~(\ref{eq:binomial})
better for increasing $M$, but the tails are not accessible.

To sample also the tails of the distribution, one does not need to raise
$M$ to $10^{18}$. Instead, one performs simulations
using biased distributions, i.e., not the original ones. This
will be explained next.

\subsection{Biased Sampling}
\label{sec:biased:sampling}

To understand the basic principle that allows us to obtain
the tails of distributions, we consider another given probability distribution
$Q$ for our system of interest. It should have nonzero
probabilities $Q(\myvec{y})>0$  for all configurations $\myvec{y}$, but
apart from this can be arbitrary in principle.
This allows us to rewrite 
Eq.~(\ref{eq:ph-average}) as \cite{rosenbluth1955}
\begin{equation}
\langle A \rangle = \sum_{\myvec{y}} A(\myvec{y}) P(\myvec{y})
=
\sum_{\myvec{y}}A(\myvec{y})\frac{P(\myvec{y})}{Q(\myvec{y})}Q(\myvec{y}) 
=
\langle AP/Q \rangle_Q \,,
\label{eq:ph-averageB}
\end{equation}
where $\langle \ldots \rangle_Q$ is the average according to $Q$.
This means we can generate numerically configurations $\myvec{y}^{(i)}$ 
with simple sampling according to a different distribution $Q$
and will arive in principle at the same result.
Since $Q$ is used instead of $P$, one says one has introduced a \emph{bias},
i.e., $Q$ is called the biased distribution.
Also, one computes a different sample mean, namely
$\{ A(\myvec{y}^i) P(\myvec{y}^i)/ Q(\myvec{y}^i)\}$ 
for this biased sampling. Eq.~(\ref{eq:ph-averageB}) tells us 
it will give still the same results, for the theoretical case
of an analytical evaluation.

Although the biased result aims in principle to estimate the same
expectation value $\langle A \rangle$,
it may be actually beneficial to use a bias.
The basic idea of the biased approach is that by suitable
choice of $Q$ one can shift the numerical sampling to a different part
of the configuration space. Thus, we could choose  that part
which is most important for the quantity $A$, e.g.,
the low-probability tails of the distribution $P$. For this reason,
biased sampling is in some fields called
\emph{importance sampling}.\footnote{But in Physics, \emph{importance sampling}
is often used to describe the sampling according to the originial, i.e.
unbiased, distribution.}

As a simple hint why this could be useful,
we consider variance reduction. The variance of the distribution
of the quantity $A$ of interest is given by 

\begin{equation}
\sigma^2(A) := 
\sum_{\myvec{y}} (A(\myvec{y})-\langle A \rangle)^2 P(\myvec{y})=
\langle A^2 \rangle - \langle A \rangle^2\,.
\label{eq:mc-var}
\end{equation}
This quantity determines, e.g., an estimate of the
error of the estimate of $\langle A \rangle$  as 
$\sigma(A)/\sqrt{(M-1)}$.\\

Now, when sampling according to $Q$, the
relevant variance is for $AP/Q$, which reads as
\begin{equation}
  \begin{split}
\sigma^2_Q(AP/Q) 
&:= 
\sum_{\myvec{y}} \biggl( \frac{A(\myvec{y})P(\myvec{y})}{Q(\myvec{y})}
-\bigg\langle \frac{AP}{Q} \bigg\rangle_Q\biggr)^2 Q(\myvec{y})\\
&=
\sum_{\myvec{y}} \biggl( \frac{A(\myvec{y})P(\myvec{y})}{Q(\myvec{y})}
-\langle A \rangle\biggr)^2 Q(\myvec{y})\, .
\label{eq:mc-varB}    
  \end{split}
\end{equation}
Now, we want to use this equation to design a suitable distribution
$Q$. As extreme artificial case, we
assume that $\langle A \rangle$ is known and $A(\myvec{y})\ge 0$
for all possible configurations $\myvec{y}$. We choose
\begin{equation}
Q(\myvec{y})=\frac{A(\myvec{y})P(\myvec{y})}{\langle A \rangle}\,.
\label{eq:mc-choiceBest}
\end{equation}
When inserting into Eq.~(\ref{eq:mc-varB}) one obtains $\sigma^2_Q(AP/Q)=0$.
This means the measurement is arbitrarily accurate! At first sight surprising,
this actually holds trivially
 because in each run of the numerical experiment the value 
$AP/Q=\langle A \rangle$ is measured, i.e., the desired result.
 Clearly, this is an rather artificial example,
 because $\langle A \rangle$ is usually  not
 known. If it was known, one would not have to perform the numerical experiment.

 Nevertheless, we can learn a bit from this example.
 First, Eq.~(\ref{eq:mc-choiceBest}) tells us that the biased sampling distribution
 will usually depend on the quantity of interest.
 This also means,
 if we are interested in several quantities for one model, we have
 to perform several independent simulations with different biases to
 obtain the full result. If we used only simple, i.e. unbiased
 sampling, only one set
 of runs is needed, where jointly all quantities of interest can be obtained.

 Second,
 Eq.~(\ref{eq:mc-choiceBest}), since $A$ appears as a multiplicative factor,
 teaches us that it is favorable to sample  more frequently
 where the measured value
 is relatively large. This makes sense, because a larger measured value will
 contribute more to the average. For the case of
 $A$ measuring a histogram bin as in Eq.~(\ref{eq:histogram_entry}), it
 is obvious that we need a considerable amount of data that is located in
 the desired bin, to estimate the corresponding probability with
 high accuracy.

\subsection{Biased Sampling for the Bernoulli Process}
\label{sec:biased:sampling:bernoulli}

As an example where biased sampling is useful,
we aim at numerically estimating  over the full support the probabilities $P(l)$
of the number of 1's in the Bernoulli process.
Still, the parameter $\alpha$ denotes  the 
the probability of observing a 1 in a single flip.
Since we measure $l$ and because
the number of 1's is directly correlated to $\alpha$, the basic idea
for biasing is
to use the original distribution, but for other values $\beta\neq \alpha$
of the
parameter. Thus, for the biased sampling probability $Q(\myvec{y})$ 
we use the   Bernoulli distribution Eq.~(\ref{eq:bernoulli}).
Therefore, we do not have to change the program at all for performing
the desired sampling. Only for the analysis according to
Eq.~(\ref{eq:ph-averageB}) we need the ratio $P/Q$ which is
\begin{eqnarray}
   P(\myvec{y})/Q(\myvec{y}) & = &
  = \frac{ \alpha^l(1-\alpha)^{n-l}}
       {\beta^l(1-\beta)^{n-l}}  \label{eq:PQ:binomial}\;.
\end{eqnarray}
Note that this ratio only depends on the quantity $l$ of interest,
thus it can be directly applied when evaluating the
histogram. There is no need to evaluate it for each single sampled 
configuration $\myvec{y}^i$.

\begin{figure}[ht]
\begin{center}
  \includegraphics[width=0.6\textwidth]{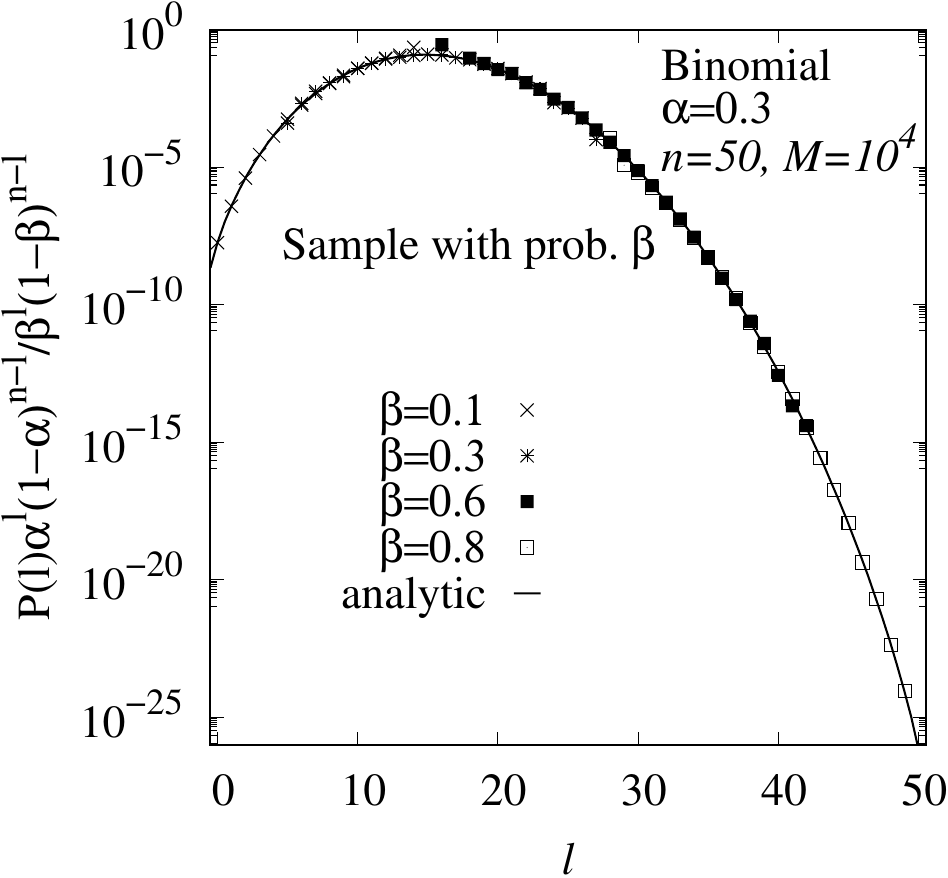}
\end{center}
  \caption{\label{fig:binomial_biased}
    Histograms of the number  of 1's obtained from  $n=50$
  coin flips for four different values $\beta$ of the probability
  to obtain a 1, rescaled to result in the distribution for $\alpha=0.3$.
  The histograms are obtained
from $M=10^4$ numerical experiments, respectively.
The solid line shows the analytic solution Eq.~(\ref{eq:binomial}).}
\end{figure}

The resulting histograms
as obtained from simulations for four different values
$\beta=0.1$, 0.3, 0.6, and 0.8, each time for $M=10^4$ samples,
and rescaled according 
to Eq.(\ref{eq:PQ:binomial}), is shown in Fig.~\ref{fig:binomial_biased}.
Note that the different histograms overlap such that
the distribution $P(l)$ is estimated over its full support,
down to probabilities as small as $10^{-25}$. A very good agreement with
the known  analytical result is visible.
In the tails of the
individual rescaled histograms the statistics becomes worse, such that here
deviations from the analytic result are visible. If one was interested
in obtaining a final
single-histogram result, one would therefore clip all raw histograms
in the low-statistics tail. For each possible value of $l$, one would use
the histogram that exhibits the best statistics.

For this example, obtaining the tails is particularly simple:
first, the quantity
of interest can be very well controlled by the  parameter $\beta$
of the biased sampling distribution $Q$, which here,
even better, is the original distribution, just for a different
value of the parameter.
Since we know how  to change the parameters
to address to region of interest, one could call this
\emph{educated sampling}.
Second, both the biasing distribution $Q$ and
the original distribution $P$ are known in detail, such that the
rescaling by multiplication with $P/Q$ is possible.

For general large-deviation simulations, it is usually hard
to relate the measurable quantity $A$
of interest to parameters of the biased sampling
distribution. A suitable name for an approach which does not need
or have this information could be \emph{blind sampling}. Note that
usually the biased sampling distribution is not known completely.
In particular, the normalization is unknown, which makes the unbiasing also more
difficult, but still possible as we will learn below.

In order to address these two problems, we first have
to be able to sample according to more or less arbitrary distributions
with possibly unknown normalization factor. This can be achieved
by using  \emph{Markov chains}, which are covered next.

\section{Markov Chain Monte Carlo Simulations}
\label{sec:MC}

We consider now a very general setup, a system with a finite number $K$ of states
$\myvec{y}=\myvec{y}_1,\myvec{y}_2,$ $\ldots,\myvec{y}_K$. The system shall be
random. Thus, we assume that the behavior of the
system is described by the probability distribution of states represented  by
probabilities  $P(\myvec{y})\ge 0$ with $\sum_{\myvec{y}}  P(\myvec{y}) =1$.

We still consider the standard target to calculate or estimate
the expectation values of observables $A(\myvec{y})$

\begin{equation}
\langle A \rangle := \sum_{\myvec{y}} A(\myvec{y}) P(\myvec{y})\,.
\label{eq:expectation}
\end{equation}

Typically, the number $K$ of states is exponentially large as
a function of some
parameter, e.g. the number  $N$ of particles or generally of degrees
of freedom. If, for example,
each degree $i$ has two states $y_{i}=0,1$ as for the Bernoulli example,
we have $K=2^N$. This means, performing a full enumeration of
Eq.~(\ref{eq:expectation}) is not feasible for even moderate values of $N$.

Often, the most simple numerical estimation procedure is 
\emph{uniform sampling}. Here one
generates a certain number  $M$ states  $\{\myvec{y}^i\}$ ($i=1,\ldots,M$) 
randomly, with uniform probability. A typical value of $M$ could be, for
example, $10^6$. For the Bernoulli example, one would obtain the uniform
sampling by simply
choosing for all entries $y_i=0,1$ with probability 0.5, instead
of the actual probabilities $(1-\alpha)$ and $\alpha$, respectively.
To estimate the
average $\langle A \rangle$, one calculates the observable for all
sampled states
and weights the results with the correct probabilities. One has to normalize
this weighted average by dividing by the sum of weights. This results in

\begin{equation}
\langle A \rangle \approx \overline{A}^{(1)} :=
\frac{\sum_{\myvec{y}^i}  A(\myvec{y}^i) P(\myvec{y}^i) }
{\sum_{\myvec{y}^j} P(\myvec{y}^j)}
\label{eq:average_uniform}
\end{equation}

It can be easily checked that this way of normalizing is correct. Say the
measurable quantity was boringly constant $A\equiv 1$,
then Eq.~(\ref{eq:average_uniform})
would result in $\sum_{\myvec{y}^i} P(\myvec{y}^i)/$
$\left(\sum_{\myvec{y}^j} P(\myvec{y}^j)\right)=1$
which is correct. Also, if the sample consisted of all $K$ possible
states, then the sum $\sum_{\myvec{y}^i} P(\myvec{y}^i)$ in the denominator
is one by normalization and Eq.~(\ref{eq:average_uniform}) reduces
to the exact expectation value Eq.~(\ref{eq:expectation}). Note also that
for Eq.~(\ref{eq:average_uniform}) it is sufficient to know the
probabilities without normalization, since this cancels from the ratio.

Unfortunately, the behavior of systems of interest is typically concentrated around
some exponentially small fraction
of states. This means that $P(\myvec{y})$ is significantly large
only for this fraction.
If, as an arbitrary example, the number of relevant states grows
exponentially like $2^{N/2}$, but the number of states like $2^{N}$,
the fraction of relevant states would decrease as $2^{N/2}/2^{N}=2^{-N/2}$.
A physical example is an Ising ferromagnet at low temperature. Here the
number of states is indeed $2^N$ but at low temperatures most states
have a finite magnetization, i.e., most spins have the same sign. Here
the fraction of relevant states is even smaller than $2^{-N/2}$. Physically
speaking, a uniformly sampled random state would basically never exhibit
a significant magnetization.

Thus, with uniform sampling one would almost sure obtain states $\myvec{y}^i$
where $P(\myvec{y}^i)$ is exponentially smaller than the probability
of relevant states, which means one misses the main contributions
to the estimation of the average
and $\overline{A}^{(1)}$ is not very accurate.

A much better approach is to  generate   $M$ configurations 
$\myvec{y}^i$ according to the desired probabilities $P(\myvec{y}^i)$,
i.e., direct sampling, as already introduced in Sec.~\ref{sec:simple:sampling}.
In this case, the sampling ensures  correct statistics and the
estimation of the expectation value $\langle A \rangle$ reduces
to the empirical average

\begin{equation}
\langle A \rangle \approx \overline{A}^{(2)} := \sum_{\myvec{y}^i} 
A(\myvec{y}^i)/M \quad (\text{with } \myvec{y}^i \sim P(\myvec{y}^i) )\,.
\label{eq:average}
\end{equation}

Note that $P$ denoted so far  the original probabilities of the system.
But, in the present context, it can be the
biased probabilities $Q$ to drive the system to desired but unlikely
configurations, as done for large-deviation sampling in section
\ref{sec:biased:sampling}. In this case, one would
have to  chose $A$ appropriately to unbias the result, as discussed
previously. We will come back to large-deviation sampling later 
in section \ref{sec:biased-mc}, but for the moment we assume just arbitrary given
probabilities $P$ according which we desire to sample,
let them be the original or biased ones.

A direct sampling can be achieved for very simple models, often by
the inversion method. This means, one can write a function in the computer
code which
upon each call returns a statistically independent
configuration according to the desired statistics.
This is indeed possible for the Bernoulli process,
where each coin flip can be simply chosen randomly.
Unfortunately, for most systems of interest, in particular if
one applies biases which depend on complex measurable quantities,
  no general approach is known, where
  one can perform this direct sampling as to obtain one
  independently drawn configuration
  from  each function call.

A general solution for this sampling problem is provided by
the Markov chain Monte Carlo
approach, which is introduced in the next section. As we will see,
it will indeed allow for sampling according to arbitrary distributions,
at least in principle. But
this comes at the price that correlations between the sampled
configurations are generated, such that only some, possible very small, fraction
of the sampled states are  statistically independent.

\subsection{Markov Chains}
\label{sec:markov}

When defining \emph{Markov chain} one  considers a set $\mathcal{S}$
of states of some elements.  The set can be
 finite, countable, or uncountable. For our applications,
i.e. in the following, we are interested in 
states that are configurations $\myvec{y}\in \mathcal{S}$
of a system described by the values of several variables
$\myvec{y}=(y_1,\ldots, y_n)$, e.g. the result of $n$ coin flips.
 
Now, a Markov chain 
is a sequence
$\myvec{y}(0)\to \myvec{y}(1) \to \myvec{y}(2) \to \ldots$
of states  $\myvec{y}(t)$   at (here) discrete times 
$t=0,1,2,\ldots$. The sequence is generated by a 
{\em probabilistic dynamic}, as explained now. The
main property of Markov chains is that state $\myvec{y}(t+1)$
depends in a stochastic way only on its preceding state
$\myvec{y}(t)$, but not on states earlier in the chain.

In case of a countable number of possible states, one
formally describes the  transitions $\myvec{y}(t)\to \myvec{y}(t+1)$
by a \emph{transition matrix}
$W_{\myvec{y}\myvec{z}}\equiv W(\myvec{y}\to\myvec{z})$ which states the 
probability\footnote{One can also define Markov chains in continuous time.
Then one would use transition rates instead of transition probabilities.}
to move from state $\myvec{y}$ (at time
$t$) to state  $\myvec{z}$ (at time  $t+1$). Usually, one considers the
case that
$W_{\myvec{y}\myvec{z}}$ does not depend on time.\\

Being probabilities, the following simple properties of
$W_{\myvec{y}\myvec{z}}$ arise:

\begin{eqnarray}
W_{\myvec{y}\myvec{z}} & \ge & 0 \quad \forall \myvec{y},\myvec{z} \in \mathcal{S}
\quad (\mbox{\it positivity})\nonumber\\
\sum_{\myvec{z}} W_{\myvec{y}\myvec{z}} & = & 1 
\quad \forall \myvec{y} \in \mathcal{S} \quad (\mbox{\it conservation})
\label{eq:W:properties}
\end{eqnarray}

The combination of state space and the transition probabilities
is called a {\em   Markov process}. When simulating a Markov process
on a computer, one speaks of a \emph{Markov chain Monte Carlo} (MCMC)
simulation.

To understand better what happens when a Markov chain is created
and to investigate the dynamics of generating, or ``running'' a Markov
chain, we assume that one does not run not only one but
a certain number $N_{\rm tot}\gg 1$
of independent Markov chains.
Then one records $N(\myvec{y},t)$, which is here
the number of chains that are in state $\myvec{y}$, at time step $t$.

\begin{example}{Two state system}
  The most simple system consists of two states, here called A and B with four
  transition probabilities. Here we assume some arbitrary values
  $W_{AA}=0.6$, $W_{AB}=0.4$, $W_{BA}=0.1$, $W_{BB}=0.9$. Thus,
  the probability to stay
  in state A is 0.6, while the probability to leave from A to B is 0.4.

  Here we assume that we run $N_{\rm tot}=100$ Markov chains
  that all start in state $A$.  
  Therefore $N(A,0)=100$
  and $N(B,0)=0$. How could the dynamics look like for these 100 chains?

  In the first transition step, about
  $N(A,0)W_{AB}=100\times 0.4=40$ chains may move from  $A\to B$,
  while the other chains remain in A and no transition $B\to A$ happens
  at $t=0$, since no chain is in state
  B.  This can be depicted as:

  $$t=0: \quad \fbox{N(A)=100} \begin{array}{c}\stackrel{0}{\longleftarrow} \\
                                     \stackrel{40}{\longrightarrow}
            \end{array} \fbox{N(B)=0}$$

  Note that the number of transitions $A\to B$ is a random quantity, so
  it may deviate substantially from 40 for 100 chains. But for the purpose
  of simplicity, we work with typical values here.

  Now, at $t=1$ we have $N(A,1)=60$ and $N(B,1)=40$. Therefore,
  $N(A,1)W_{AB}=$ $60\times 0.4=24$ chains may move from  $A\to B$,
  while $N(B,1)W_{BA}=40\times 0.1=4$ may move from $B\to A$, while
  the other chains exhibit no change of state:

  $$t=1: \quad \fbox{N(A)=60} \begin{array}{c}\stackrel{4}{\longleftarrow} \\
                                     \stackrel{24}{\longrightarrow}
            \end{array} \fbox{N(B)=40}$$

  Now, at $t=2$ we have $N(A,2)=40$ and $N(B,2)=60$. The subsequent dynamics until
  $t=7$, again when considering only the most likely evolution, may look
  as follows:
  
  $$t=2: \quad \fbox{N(A)=40} \begin{array}{c}\stackrel{6}{\longleftarrow} \\
                                     \stackrel{16}{\longrightarrow}
            \end{array} \fbox{N(B)=60}$$

    $$t=3: \quad \fbox{N(A)=30} \begin{array}{c}\stackrel{7}{\longleftarrow} \\
                                     \stackrel{12}{\longrightarrow}
            \end{array} \fbox{N(B)=70}$$

    $$t=4: \quad \fbox{N(A)=25} \begin{array}{c}\stackrel{8}{\longleftarrow} \\
                                     \stackrel{10}{\longrightarrow}
            \end{array} \fbox{N(B)=75}$$

    $$t=5: \quad \fbox{N(A)=23} \begin{array}{c}\stackrel{8}{\longleftarrow} \\
                                     \stackrel{9}{\longrightarrow}
            \end{array} \fbox{N(B)=77}$$

    $$t=6: \quad \fbox{N(A)=22} \begin{array}{c}\stackrel{8}{\longleftarrow} \\
                                     \stackrel{9}{\longrightarrow}
            \end{array} \fbox{N(B)=78}$$

    $$t=7: \quad \fbox{N(A)=21} \begin{array}{c}\stackrel{8}{\longleftarrow} \\
                                     \stackrel{8}{\longrightarrow}
            \end{array} \fbox{N(B)=79}$$

  From now on one has about $N(A,t)W_{AB}=N(B,t)W_{BA}$. This means, the transfer
  between the two states balances leading to $N(Y,t)=$const for $Y=A$
  and $Y=B$.
  Thus, a \emph{stationary state} is reached and $N(Y,t)=N(Y)$ will
  become time-independent when ignoring relatively small
  fluctuations, at least if the number $N_{\rm tot}$ of states
  is very large. 
\end{example}

To allow for a formal description of the dynamics of a Markov chain,
we define $P(\myvec{y},t)$ $=\lim_{N_{\rm tot}\to \infty}
\langle N(\myvec{y},t)/N_{\rm tot} \rangle $ be
the probability that the system at time $t$ is in state
$\myvec{y}(t)=\myvec{y}$. 

\begin{figure}[ht]
\begin{center}
{\scalebox{0.6}{\includegraphics{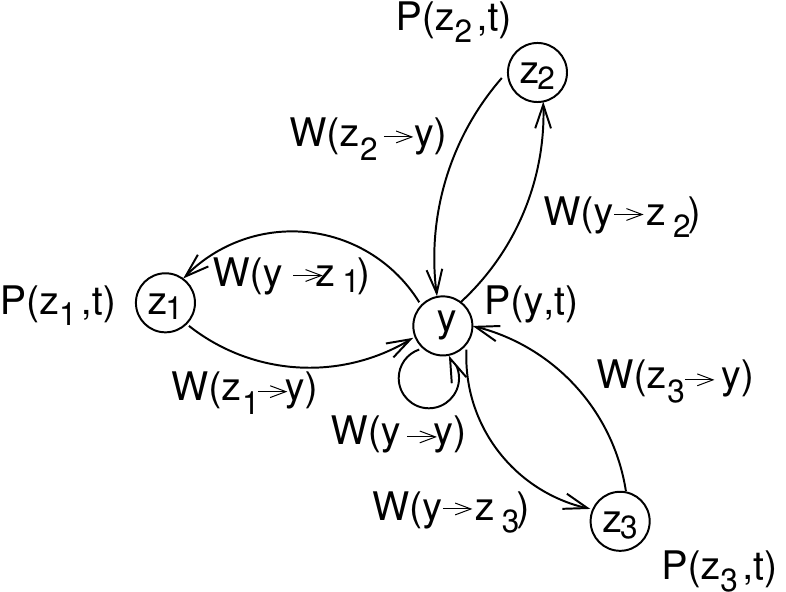}}}
\end{center}
\caption{\label{fig:master:equation}
  Flow of probability from and to state $y$. This and other states
  are shown as circles.
  It is assumed that $y$  has non-zero transition probabilities 
  to and from three other states $z_1$, $z_2$ and $z_3$,
  shown as arrows. Other possible
  transitions of the states $z_1$, $z_2$ and $z_3$ are not shown here.}
\end{figure}

We want to write down a balance equation
for the probability $P(\myvec{y},t)$ to be in  state $\myvec{y}$.
Usually, this fundamental equation is
called \emph{Master Equation}. State $\myvec{y}$ will have connections,
i.e.\ non-zero transition probabilities to other states, see
Fig.~\ref{fig:master:equation}.  This may lead
to some chains move into state $\myvec{y}$, i.e.\ a flow of probability
into state $\myvec{y}$, which results in a contribution increasing
$P(\myvec{y},t+1)$ with respect to $P(\myvec{y},t)$.
This flow of probability will be proportional to the probability
being in another state $\myvec{z}$ and proportional to the
transition probability $W_{\myvec{z}\myvec{y}}$, thus proportional to
$W_{\myvec{z}\myvec{y}}P(\myvec{z},t)$.
On the other
hand, some chains will be in state $\myvec{y}$ and move out of other states
$\myvec{z}$. Hence, this flow of probability is given by 
$W_{\myvec{y}\myvec{z}}P(\myvec{y},t)$ and it contributes negatively
to $P(\myvec{y},t+1)$.

This leads to the following equation for the change 
$\Delta P(\myvec{y},t)$ of the probability from time $t$ to $t+1$:

\begin{equation}
\Delta P(\myvec{y},t) := P(\myvec{y},t+1)-P(\myvec{y},t) = \sum_{\myvec{z}} 
W_{\myvec{z}\myvec{y}}P(\myvec{z},t) - 
\sum_{\myvec{z}} W_{\myvec{y}\myvec{z}}P(\myvec{y},t) \quad \forall \myvec{y}\in \mathcal{S}\,.
\label{eq:master}
\end{equation}

Under specific conditions \cite{reichl1998}, as specified by the Theorems
of Perron \cite{perron1907} and Frobenius \cite{frobenius1912},
the probability distribution will converge to
a  stationary, i.e.\ 
time-independent distribution

\begin{equation}
  P_{\rm st}(\myvec{y}) := \lim_{t\to\infty} P(\myvec{y},t)\,.
  \label{eq:stationary}
\end{equation}
This will happen in particular if largest left eigenvalue
$\lambda=1$ of the matrix
$W$ has multiplicity 1, i.e., the corresponding eigenspace is one-dimensional.

\begin{figure}[ht]
\begin{center}
{\scalebox{0.6}{\includegraphics{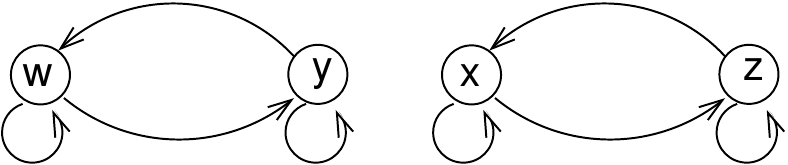}}}
\end{center}
\caption{\label{fig:non-ergoic}
  Example for a non ergodic system. Possible transitions
  with non-zero transition probabilities are indicated by arrows.
  A Markov chain which starts in $y$=W
  or $y=$ Y will always have $P$(X,$t$)$=0$ and $P$(Z,$t$)$=0$ while
  this will not be the case if the chain started in X or Z. Thus,
  the limiting distribution will depend on the initial state.
 }
\end{figure}

The limit Eq.~(\ref{eq:stationary}) has to be
independent of any given initial state $\myvec{y}(0)$. The
Markov process is called {\em ergodic} in this case. This basically means that
one can reach each state from any other state by a finite
sequence of transitions.
Note that there are non-ergodic systems, a simple example is shown in
Fig.~\ref{fig:non-ergoic}.

Now, after having introduced Markov chains in a general way, we come
back to the target of sampling configurations $\myvec{y}$ of a system 
according to  given probabilities $P(\myvec{y})$. For a Markov chain
that achieves a stationary probability distribution $P_{\rm st}$
this can be achieved by

\begin{center}
\fbox{choose $W_{\myvec{y}\myvec{z}}$ such that $P_{\rm st}=P$.}
\end{center}

Since $P_{\rm st}=P$ is time-independent we simply
replace $P(\myvec{y},t)$ by $P(\myvec{y})$. This implies 
from Eq.~(\ref{eq:master}) that

$$
0=\Delta P(\myvec{y})= \sum_{\myvec{z}} W_{\myvec{z}
\myvec{y}}P(\myvec{z}) - 
\sum_{\myvec{z}} W_{\myvec{y}\myvec{z}}P(\myvec{y}) \quad
\forall \myvec{y}\in \mathcal{S}\,.
$$
This means one has obtained more conditions for the transition probabilities,
in addition to Eqs.~(\ref{eq:W:properties}), depending on
the target probabilities $P(\myvec{y})$. These conditions are called
\emph{global balance}, because for each state the total inflow of probability,
globally summed over all other states,
balances with the total outflow. In particular, the
global balance conditions relate the transition probabilities to the
desired sampling probabilities $P(\myvec{y})$.

A very convenient way to fulfill the global balance is to make 
each pair of corresponding contributions of the two sums cancelling
each other, i.e., 
\begin{equation}
 W_{\myvec{z}\myvec{y}}P(\myvec{z}) - W_{\myvec{y}\myvec{z}}
P(\myvec{y})=0 \quad \forall \myvec{y},\myvec{z}\in \mathcal{S}
\label{eq:detailed_balance}
\end{equation}
This is called {\em detailed balance}, because the cancelation
of probability flow appears for each pair of states. These
conditions help a great deal in setting up suitable transition
probabilities that lead to Markov chains
generating configurations according to the target probabilities $P(\myvec{y})$.
One special widely-used algorithm, the Metropolis-Hastings
algorithm, will be presented in Sec.~\ref {sec:metropolis}.

In general and even if one takes the Metropolis-Hastings algorithm, 
there is a lot of freedom in choosing the transition probabilities,
influencing strongly the efficiency of the corresponding algorithm.
One general rule is, the more the configurations are changed
between two time steps, the faster
the algorithm will walk through configuration space,
thus the more efficient the algorithm is. Unfortunately, these
system-wide changes are hard to set up, often one has to be satisfied
with single-variable changes.

The efficiency of the Markov chain Monte Carlo approach is visible through
two properties: the speed of \emph{equilibration} and through the
occuring temporal \emph{correlations}.
Equilibration refers to the fact that the Markov chains start in some
configuration $\myvec{y}(0)$. Since typically, the configurations
are changed only a bit between two consecutive steps of the Markov chain,
it will take a certain number of steps $t_{\rm equi}$ to ``forget''
this initial configuration.
Thus, when calculating averages, the configurations obtained
for   $t<t_{\rm equi}$ have to be omitted. In the next section,
an example is shown
which illustrates how the equilibration becomes visible in a
simulation outcome, allowing one to estimate $t_{\rm equi}$.

The fact that configurations change only gradually over time,
means $\myvec{y}(t+1)$ is typically similar to
$\myvec{y}(t)$. Thus, configurations which are close in time $t$
are  correlated even after equilibration. Since one is interested
in sampling statistically independent configurations,
 only distant states $\myvec{y}(t), 
\myvec{y}(t+\Delta t),
\myvec{y}(t+2\Delta t),\dots$ should be included in the measurements.

Note that 
$t_{\rm equi}$ and $\Delta t$ depend strongly on 
the considered system, on the applied algorithm and
on the parameters used for the simulation. Usually, they
cannot be calculated in advance but
have to be determined experimentally in exploratory 
test simulations. Some systems, in particular if they are large, are
too hard to simulate such that even equilibration cannot be achieved
in reasonable time with the
algorithms at hand.

\subsection{MCMC for the Bernoulli Process}
\label{mc:bernoulli}

As a simple toy application of Markov chain Monte Carlo simulations,
we consider again the Bernoulli process, i.e., configurations
$\myvec{y}=(y_1,y_2,\ldots, y_n)$ of $n$ coin tosses $y_i\in \{0,1\}$.
Like before, we use the parameter $\alpha \in [0,1]$ to define
the probabilities for the two possible outcomes of each coin toss,
see Eq.~(\ref{eq:coin:toss}).

Now, we want to use a Markov chain to sample states according to the
configuration
probabilities Eq.~(\ref{eq:bernoulli}). Note that for this simple model,
a Markov chain is actually not needed, since one can simulate the coin tosses
efficiently by directly sampling, as done in Sec.~\ref{sec:simple:Bernoulli}.
Thus, this simple application is for pedagogical reasons only.

As mentioned before, there are in general many possibilities
to set up the transition probabilities in the matrix  $W$.
Here we proceed as follows: Given the current state
$\myvec{y}=\myvec{y}(t)$, we just select
$n_{\rm c}$ times a randomly chosen
entry and redraw it according to the original probabilities. Therefore,
 we perform $n_{\rm c}$ new coin tosses, while the outcomes
of the other coin tosses contained in $\myvec{y}$ are not touched. This
is implemented by the following algorithm:

\begin{algorithm}{Bernoulli-MC($\myvec{y}$, $n$, $n_{\rm c}$)}
  \> {\bf do} $n_{\rm c}$ times:\\
  \> {\bf begin}\\
  \> \> choose random index $i\in \{1,\ldots,n\}$\\
  \>\> redraw $y_i$ according to (\ref{eq:coin:toss})\\
  \> {\bf end}\\

\end{algorithm}

The resulting configuration  is $\myvec{y}(t+1)$.
Each time a new index $i$ is drawn
randomly and independently, this might result in repeated
redrawing some entries.  Since each coin toss is independent,
only the last one will be effective for a given entry.
Thus, the actual number of entries actually redrawn may be smaller than
$n_{\rm c}$.
An implementation in C is contained \cite{bernoulli_code2024}
in {\tt mc\_bernoulli.c}\footnote{The
code can be compiled in a Unix shell by
{\tt cc -o mc\_bernoulli mc\_bernoulli.c -lm} .} 
in the function {\tt bernoulli\_mc\_step0()}.

It is intuitively clear that changing some entries according to
the original probabilities, while leaving all others untouched,
which nevertheless have previously been
drawn in this way, will lead to the correct sampling. Still, we want
to check detailed balance explicitly.
The desired sampling probability of configuration $\myvec{y}$
is given by Eq.~(\ref{eq:bernoulli}).

For obtaining the transition probabilities, we
consider two states
$\myvec{y}$ and $\myvec{z}$ that are connected by
a step in the Markov chain. 
Let $I_{\rm c}$ contain those indices where $\myvec{y}$ and $\myvec{z}$
differ, i.e., $I_{\rm c}=\{i|y_i \neq z_i; i=1,\ldots,n\}$. Note that
$\myvec{y}$ and $\myvec{z}$ can have a non-zero transition probability
only when $n_{\rm c}\ge |I_{\rm c}|$ .

Furthermore, let $I_{\rm s}$ be the actual set of indices  selected
by the algorithm, with  $|I_{\rm s}| \le n_{\rm c}$. This set
will most of the time 
be a super set  of $I_{\rm c}$, because for some of the selected entries
redoing the coin flip might lead to the previous outcome.
Hence,  for the entries in $I_{\rm s}\setminus I_{\rm c}$,
the entries are not changed, like those entries which do not appear
in $I_{\rm s}$ anyway. We denote by $R(I_{\rm s})$ the probability that
the algorithm selects a certain subset $I_{\rm s}$, which might be
a complicated function of $I_{\rm s}$. For the entries $i$ which appear
in $I_{\rm s}$, the joint probability for obtaining the
outcomes $z_i$ is simply $\prod_{i \in I_{\rm s}}\alpha^{z_i}(1-\alpha)^{1-z_i}$.
Splitting $I_{\rm s}$ into $I_{\rm c}$ and $I_{\rm s}\setminus I_{\rm c}$,
and summing over all possible sets $I_{\rm s}$, this leads to
the following transition probability:

\begin{equation*}
W_{\myvec{y}\myvec{z}}  =  \sum_{I_{\rm s}} R(I_{\rm s}) 
\prod_{i \in I_{\rm s}\setminus I_{\rm c} }\alpha^{z_i}(1-\alpha)^{1-z_i}
\prod_{i \in I_{\rm c}}\alpha^{z_i}(1-\alpha)^{1-z_i}
\end{equation*}

Since $I_{\rm c}$ is fixed, the last product factor is the same for all
terms in the sum.
Thus, it can be taken out of the sum, leading to

\begin{equation*}
  W_{\myvec{y}\myvec{z}}  =  \prod_{i \in I_{\rm c}}\alpha^{z_i}(1-\alpha)^{1-z_i}
  \left(\sum_{I_{\rm s}} R(I_{\rm s}) 
\prod_{i \in I_{\rm s}\setminus I_{\rm c} }\alpha^{z_i}(1-\alpha)^{1-z_i}\right)\,,
\end{equation*}

The transition probability for the opposite move is obtained
by replacing $z_i$ with $y_i$. The selected subsets $I_{\rm s}$
which contribute to the transition are the same as for the forward move.
This means they
have the same probabilities $R(I_{\rm s})$. Therefore, as we will see below,
that we do not have to know the function dependency of $R(I_{\rm s})$:

Let's now check detailed balance Eq.~(\ref{eq:detailed_balance}) explicitly.
For this purpose, we split the product of $P(\myvec{y})$ into
two factors, containing contributions from $I_{\rm c}$ and from
all other entries.
Also,  we use that $y_i=z_i$
for $i \not\in I_{\rm c}$, in particular for $i \in I_{\rm s}\setminus I_{\rm c}$:

\begin{eqnarray*}
P(\myvec{y})W_{\myvec{y}\myvec{z}} & = &
\underbrace{\prod_{i\in I_{\rm c}} \alpha^{y_i}(1-\alpha)^{1-y_i}}_{A}
  \underbrace{\prod_{i\neq I_{\rm c}}\alpha^{y_i}(1-\alpha)^{1-y_i}}_{B} \\
& & \times 
\underbrace{\prod_{i \in I_{\rm c}}\alpha^{z_i}(1-\alpha)^{1-z_i}}_{C}
  \underbrace{\left(\sum_{I_{\rm s}} R(I_{\rm s}) 
\prod_{i \in I_{\rm s}\setminus I_{\rm c} }\alpha^{z_i}(1-\alpha)^{1-z_i}\right)}_{D}
\\
& \stackrel{y_i=z_i; i\neq\in I_{\rm c}}{=} &
\prod_{i\in I_{\rm c}} \alpha^{z_i}(1-\alpha)^{1-z_i}
  \prod_{i\neq I_{\rm c}}\alpha^{z_i}(1-\alpha)^{1-z_i} \\
& & \times 
\prod_{i \in I_{\rm c}}\alpha^{y_i}(1-\alpha)^{1-y_i}
  \left(\sum_{I_{\rm s}} R(I_{\rm s}) 
\prod_{i \in I_{\rm s}\setminus I_{\rm c} }\alpha^{y_i}(1-\alpha)^{1-y_i}\right)
\\
& \stackrel{(*)}{=} & P(\myvec{z})W_{\myvec{z}\myvec{y}}\,,
\end{eqnarray*}
where from the second to the third expression,
the factors $A$ and $C$ were just exchanged, and for factors $B$ and $D$ 
$y_i=z_i$ for  $i\not\in I_{\rm c}$ was used. Thus, detailed balance holds.

\begin{figure}[ht]
\begin{center}
\includegraphics[width=0.55\textwidth]{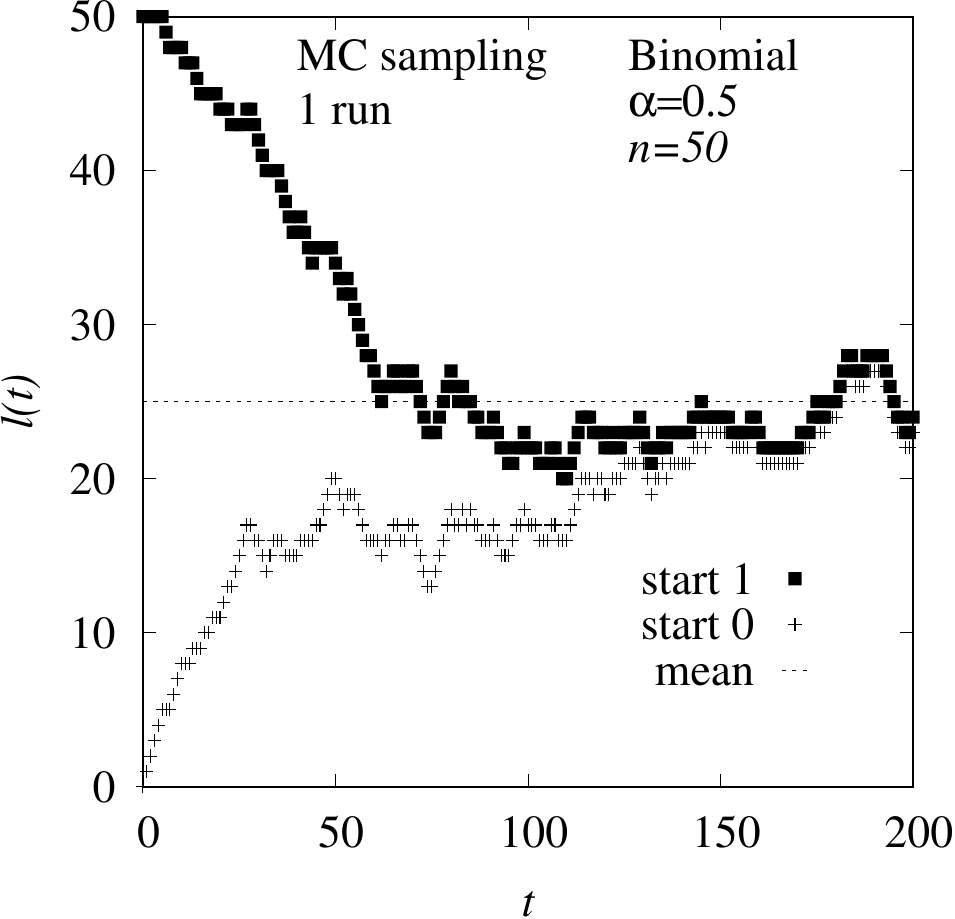}
\end{center}
\caption{Sample Markov chain for the Bernoulli case for $n=50$ coin flips:
  Number $l(t)=\sum_i y_i(t)$ of 1's as function of the Monte Carlo step $t$
  for probability $\alpha=0.5$ and $n_{\rm c}=2$. Two
extreme different initial configurations are taken into account.
The horizontal line
  indicates the expectation value $\alpha n$.
\label{fig:mc_binomial_run1}
}
\end{figure}

Now the equilibration of the Monte Carlo simulation is considered.
To investigate this, two extreme start configuration
$\myvec{y}(0)=(y^0,y^0, \ldots, y^0)$ are used,
with all entries being the same, either 
all  $y^0=0$ or all $y^0=1$.
Fig.~\ref{fig:mc_binomial_run1}
shows the  time evolution of the number $l$ of 1's as function of the Monte
Carlo step $t$ for the case of $n_{\rm c}=2$ changes per Monte Carlo step.
After about $t=150$ steps, the two simulations, which were generated
with the same set of random numbers, start fluctuating both
around the expectation value $\alpha n=25$.
Thus, the equilibration time $t_{\rm equi}$ is about 150 steps.

\begin{figure}[htb]
\begin{center}
\includegraphics[width=0.55\textwidth]{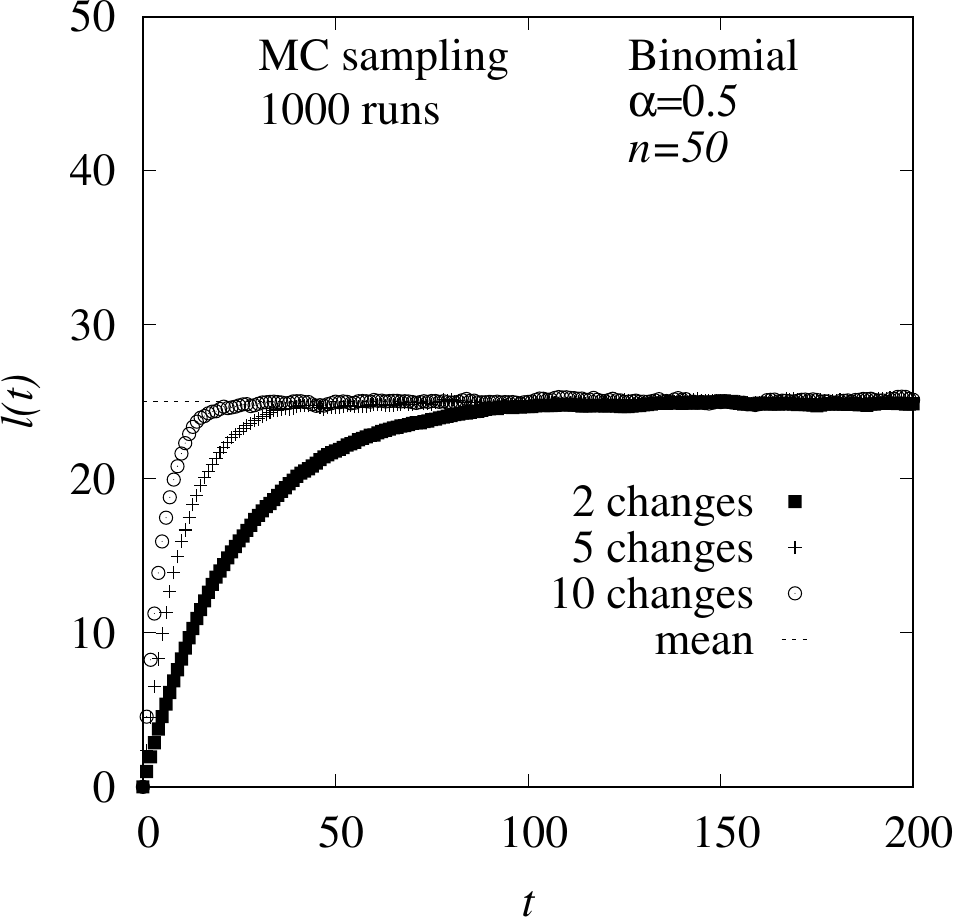}
\end{center}
\caption{Markov chain for the Bernoulli case with $n=50$ coin flips
  and $\alpha=0.5$. Shown is the
  number $l(t)=\sum_i y_i(t)$ of 1's, averaged over 1000 runs,
  starting at $l(0)=0$. Three different  numbers $n_{\rm c}=1,2,5,10$
  of entry changes are considered. The horizontal line
  indicates the expectation value $\alpha n$.
\label{fig:mc_binomial_run2}
}
\end{figure}

In Fig.~\ref{fig:mc_binomial_run2} $l(t)$ is shown for different
number $n_{\rm c}$ of selected entries for changes. The more entries
are changed, the faster the convergence is. This is in general true
for all MCMC algorithms: the approach is the more
efficient, the faster it can move through configuration space.

\subsection{Metropolis-Hastings algorithm}
\label{sec:metropolis}

Often it is not possible to set up the
transition probabilities $W_{\myvec{y}\myvec{z}}$  directly. This is typically
the case for more realistic
models or more complex distributions. In particular, when we use
large-deviation
sampling where the bias depends on the measured quantity of interest, this
will occur, as explained in Sec.~\ref{sec:biased-mc}.

For such cases,
the \emph{Metropolis-Hastings} algorithm \cite{metropolis1953,hastings1970}
provides a general framework, which makes it way simpler to obtain
correct transition probabilities for general problems. 
The algorithm works well even for the case, where
the normalization constant of the target
probabilities $P(\myvec{y})$ is unknown.

The basic idea works as follow. We assume that we are
given the current configuration $\myvec{y}=\myvec{y}(t)$ of
the Markov chain. A Monte Carlo step consists of two parts, which are defined
by \emph{two} matrices  $A(\myvec{y}\to \myvec{z})$ and
$\tilde{W}(\myvec{y}\to \myvec{z})$ of probabilities:

\begin{enumerate}
\item Select a \emph{trial configuration} $\myvec{z}$ {\em randomly},
  defined by  an ``arbitrarily'' chosen matrix
  $A(\myvec{y}\to \myvec{z})$.
\item With probability  $\tilde{W}(\myvec{y}\to \myvec{z})$, 
  the configuration $\myvec{z}$ is {\em accepted}, i.e.,
  $\myvec{y}(t+1)=\myvec{z}$.
With probability $1-\tilde{W}(\myvec{y}\to \myvec{z})$, 
configuration $\myvec{z}$ is {\em rejected}, i.e.,
$\myvec{y}(t+1)=\myvec{y}$
\end{enumerate}

Note that the matrix $A(\myvec{y}\to \myvec{z})$ is not to be confused
with the measurable quantity $A(\myvec{y})$.
$\tilde{W}(\myvec{y}\to \myvec{z})$ is called {\em acceptance probability}.
At first sight, it might seem that one has not gained anything, because
instead of selecting the transition probabilities
$W_{\myvec{y} \myvec{z}}=W(\myvec{y}\to \myvec{z})$ one has to choose
two probability matrices $A(\myvec{y}\to \myvec{z})$  and
$\tilde{W}(\myvec{y}\to \myvec{z})$ . The main advantage of this approach
is, as we will see soon, that one is basically free to choose $A$ and then 
$\tilde{W}$ is set up such that the correct sampling according to the
desired target probabilities $P(\myvec{y})$ is ensured.

To understand why this works, we first realize that the two-step
process makes the total transition probability a product of the
probabilities of the two steps.

\begin{equation}
W(\myvec{y}\to \myvec{z}) = 
A(\myvec{y}\to \myvec{z})\tilde{W}(\myvec{y}\to \myvec{z})
\quad (\myvec{y}\neq \myvec{z})\,.
\label{eq:rewrite}
\end{equation}

This means, the total probability to stay in  $\myvec{y}$ is
given by 
$W(\myvec{y}\to \myvec{y})=
1-\sum_{\myvec{z}\neq \myvec{y}}W(\myvec{y}\to \myvec{z})$.

Now, we insert Eq.~(\ref{eq:rewrite}) into the detailed balance condition 
Eq.\ (\ref{eq:detailed_balance}). We assume that $A$ is somehow given and
resolves with respect to the unknown quantities $\tilde W$ and obtain
\begin{equation}
\frac{\tilde{W}(\myvec{y}\to \myvec{z})}{\tilde{W}(\myvec{z}\to \myvec{y})}
= \frac{P(\myvec{z})}{P(\myvec{y})}
\frac{A(\myvec{z}\to \myvec{y})}{A(\myvec{y}\to \myvec{z})}\,.
\label{eq:detailed_balanceB}
\end{equation}

Now we have to select some acceptance probability matrix $\tilde W$
that fulfills this equation. For the Metropolis-Hastings algorithm
\cite{metropolis1953,hastings1970} the choice is
\begin{equation}
\tilde{W}(\myvec{y}\to \myvec{z}) = 
\min \Bigg(1, \underbrace{\frac{P(\myvec{z})}{P(\myvec{y})}
\frac{A(\myvec{z}\to \myvec{y})}{A(\myvec{y}\to \myvec{z})}}_{=:Q} \Bigg)\,,
\label{eq:metropolis}
\end{equation}

Note that for the opposite move $\myvec{z}\to \myvec{y}$
the vectors $\myvec{z}$ and $\myvec{y}$ exchange and therefore
$Q$ is replaced by $1/Q$. This also means that for one of the two moves
between $\myvec{z}$ and $\myvec{y}$ the transition probability is one.
One could use smaller acceptance probabilities, e.g. $1/2$ instead
of 1 for one direction, the first term of the $\min\{\ldots \}$.
But then one would have, to
conserve detailed balance, to use the half of
the probabilities for the opposite direction as well resulting in 
$\tilde{W}(\myvec{y}\to \myvec{z}) = 
\min \Bigg(\frac 1 2, \frac 1 2\frac{P(\myvec{z})}{P(\myvec{y})}
\frac{A(\myvec{z}\to \myvec{y})}{A(\myvec{y}\to \myvec{z})} \Bigg)
$,
but this would reduce the overall number of performed changes.
On the other hand,
it is not possible to increase the acceptance rate 1 for the first
term in the $\min\{\ldots \}$.
In this sense, the Metropolis-Hastings choice is maximal with respect
to the number of accepted moves, given the trial
construction probabilities $A(\myvec{z}\to \myvec{y})$.

We can now verify  that the rewritten detailed balance condition
 holds.
Without loss of generality we assume for the ratio  $Q<1$. Then we have 

\begin{eqnarray*}
\tilde{W}(\myvec{y}\to \myvec{z}) &= & Q\\
\tilde{W}(\myvec{z}\to \myvec{y}) & = & 1 \\
\Rightarrow
\frac{\tilde{W}(\myvec{y}\to \myvec{z})}{\tilde{W}(\myvec{z}\to \myvec{y})}
& = & Q/1 =Q\,,
\end{eqnarray*}

as required by Eq.~(\ref{eq:detailed_balanceB}).
The space of possible algorithms is represented by the trial construction
probabilities $A(\myvec{y}\to \myvec{z})$, thus it is very large.
For sure, these values have to be selected such that the resulting
Markov process is ergodic.
 Thus, the probabilities have to be set up in a way such that
between any pair of states $\myvec{y}$ and $\myvec{z}$ there
is either a direct transition possible with $A(\myvec{y}\to \myvec{z})>0$,
or there exists a finite sequence $\myvec{x_1},\ldots,\myvec{x_k}$
of intermediate states such that $A(\myvec{y}\to \myvec{x_1})>0$,
$A(\myvec{x_i}\to \myvec{x_{i+1}})>0$ for $i=1,\ldots,k-1$
and $A(\myvec{x_{k}}\to \myvec{z})>0$.

Typically, the generation of the trial configuration is implemented
by changing one or several randomly chosen entries with respect to the 
current configuration $\myvec{y}(t)$.
In general, the more entries are changed, 
while keeping the acceptance probability $\tilde{W}$
high, the faster the algorithm.
Unfortunately, there is usually a trade off between acceptance
probability and how much
the trial state configuration differs from the current configuration.
The most simple and often used approach is to change just one entry
of the current configuration to obtain the trial one. This is
done in the example in the following section.

\subsection{Metropolis-Hastings Algorithm for the Bernoulli Process}

Again, we consider as simplest toy example the
Bernoulli process for $n$ coin tosses $\myvec{y}\in \{0,1\}^n$ with probability
$\alpha$  for obtaining a 1, see Eq.~(\ref{eq:coin:toss}),
described by $P(\myvec{y})$
according to Eq.~(\ref{eq:bernoulli}).

Let $\myvec{y}=\myvec{y}(t)$ be the current configuration of a Markov chain
at step $t$.
A very simple \emph{choice} for the generation of a trial
configuration $\myvec{z}$
is:

\begin{itemize}
\item Choose one entry $i_0 \in \{1,\ldots,n\}$ with uniform probability $1/n$.

\item  
 \begin{equation}
\text{Let} \quad   z_i= \left\{
  \begin{array}{ll} 1-y_i & i=i_0 \\
    y_i & i\neq i_0
  \end{array} \right.\,.
  \label{eq:bernoulli_single}
\end{equation}
\end{itemize}

Since exactly one entry is changed with respect to $\myvec{z}$,
this is called a \emph{single variable flip} algorithm. 
In statistical physics, e.g., 
for Ising spin systems like ferromagnets or spin glasses,
it is called the \emph{single spin flip} algorithm.

The corresponding trial construction matrix is
$A(\myvec{y}\to \myvec{z}) = A(\myvec{z}\to \myvec{y}) = 1/n$, if
$\myvec{y}$ and $\myvec{z}$ differ by exactly one entry. For all other
cases we have
$A(\myvec{y}\to \myvec{z}) = A(\myvec{z}\to \myvec{y}) = 0$.
For the latter case, no transitions occur between the pairs of
configurations, so detailed balance holds
trivially. We only have to consider the case where
$\myvec{y}$ and  $\myvec{z}$ differ by one entry.

The calculation of the Metropolis
acceptance probability Eq.~(\ref{eq:metropolis}), using
Eq.~(\ref{eq:bernoulli_single}) yields

\begin{eqnarray*}
  \tilde{W}(\myvec{y}\to \myvec{z})  & = &
\min\left(1,   \frac{\prod_{i=1}^n\alpha^{z_i}(1-\alpha)^{1-z_i}}
         {\prod_{i=1}^n\alpha^{y_i}(1-\alpha)^{1-y_i}}
\frac{1/n}{1/n} \right)\\
       & = & \min\left(1,\frac{\alpha^{z_{i_0}}(1-\alpha)^{1-z_{i_0}}}
       {\alpha^{y_{i_0}}(1-\alpha)^{1-y_{i_0}}}\right)\\
       & = & \min\left(1,\frac{\alpha^{1-y_{i_0}}(1-\alpha)^{y_{i_0}}}
       {\alpha^{y_{i_0}}(1-\alpha)^{1-y_{i_0}}}\right) \\
       & = & \min\left(1,\alpha^{1-2y_{i_0}}(1-\alpha)^{2y_{i_0}-1} \right)\,
\end{eqnarray*}

Hence, one can calculate the acceptance probability without
performing the actual flip beforehand. Thus, one performs a single-variable
flip with probability $\tilde W$ to yield $\myvec{y}(t+1)$,
otherwise the current configuration $\myvec{y}(t)$ is kept.
A corresponding C code \cite{bernoulli_code2024}
is contained in {\tt mc\_bernoulli.c}
in the function {\tt bernoulli\_metropolis()}.

The advantage of changing only a single variable is that such
an approach works for most systems, in particular complex systems, where
no direct sampling is possible.
Still, since at most one variable is flipped in each Monte Carlo step,
the algorithm is rather slow. In principle it can be
made faster  if more than one variable is changed
within on step, yielding faster equilibration and
a smaller correlation time.
But if too many changes are performed, the acceptance rate becomes lower.

\section{Biased Markov Chain Sampling}
\label{sec:biased-mc}

Now we come back to the main target, to obtain for 
a random
process the distribution $Q(S)$ of
a measurable scalar quantity $S$, which we also call \emph{score}.
The score can be in principle any scalar function of the entire configuration
$\myvec{y}$.
The distribution can be obtained in principle by considering
all possible configurations $\myvec{y}$ and adding up the probabilities
$P({\myvec{y}} )$ of those configurations for which
$S(\myvec{y})$ has the desired value $S$, i.e.

\begin{equation}
  Q(S) = \sum_{\myvec{y}} \delta_{S,S(\myvec{y})} P({\myvec{y}} )\,.
  \label{eq:def:Q:S}
\end{equation}
We aim at obtaining
$Q(S)$ over a large range of the support, down to the tails,
where the probabilities are very small like $10^{-50}$ or even lower.

Again, like in Sec.~\ref{sec:biased:sampling}, we introduce a \emph{bias}
that changes the sampling distribution such that it is shifted to
the region of interest. For  the previous approach, when applied to the
Bernoulli process, 
we knew that by using a different coin-flip probability $\beta\neq \alpha$
we could shift the distribution $P(l)$ of the number $l$ of 1's
to the tails. We called this \emph{educated} sampling, since we know how we
have to change the natural parameter of the distribution to concentrate
the sampling in the region of interest.
In contrast to this, we will now consider the
more general case, where we have no idea of how we could influence
the sampling by just changing some parameters of the system.
Often, as we will show in an example below, there is even no suitable
connection of $S$ to system parameters.
This more general approach, which allows one to sample the
region of interest without prior knowledge, we call \emph{blind sampling}.

The basic idea of this blind sampling
is to make the bias depend on the quantity $S$
of interest and some control parameter. This
will drive the biased simulation
automatically in the desired  region, but we do not have to know
how the system has to be controlled to do this.

\subsection{Exponential Bias}

A standard approach, and for some applications even the optimal one,
is to use an \emph{exponential bias},
which for a  configuration $\myvec{y}$ reads
\begin{equation}
B_{\Theta}(\myvec{y})= \exp(-{S}(\myvec{y})/\Theta)\,,
\label{eq:canonical}
\end{equation}
where ${S}(\myvec{y})$ is the score evaluated for the configuration. 
This bias 
appears very natural for physicists since it corresponds to the Boltzmann
weight of the \emph{canonical ensemble}. Therefore
$\Theta$ is a temperature-like parameter, short just temperature, which controls
the impact of the bias. Note that also other biases are possible,
in principle everything that works is good. Nevertheless, for many
systems the exponential bias works very well and it is also used 
for analytical calculations, where one can take advantage
of biases to obtain tails of distributions
in a similar way. For a short introduction, see Ref.\ \cite{majumdar2017}.
Also more detailed texts, still well readable for physicists,
are available \cite{touchette2009,touchette2011}.

We consider the case that the full distribution of configurations is
obtained by the original probability $P(\myvec{y})$ times the bias.
Like any distribution, 
it should be normalized, leading to  the probability 
$P_\Theta(\myvec{y})$ to obtain a configuration $\myvec{y}$ in
the biased ensemble as given by

  \begin{equation}
    P_\Theta(\myvec{y})=\frac{1}{Z(\Theta)} P(\myvec{y}) B_{\Theta}(\myvec{y})\;,
    \label{eq:biased:target:distr}
  \end{equation}
with the normalization $Z=\sum_{\myvec{x}}P(\myvec{x}) B_\Theta(\myvec{x})$.

From physical intuition, it is easy to understand the effect of the
bias depending on the temperature $\Theta$:
\begin{itemize}
\item For $\Theta=\infty$ the argument of the exponential is zero.
  Thus, the bias has no effect and one is back to standard sampling
  according to $P(\myvec{y})$. Thus with respect to $Q(S)$,
  typical values will be  sampled, in the region where the
  probability $Q(S)$ is large.
  
\item If $\Theta \to 0^+$, the temperature is small but positive. This
  means, from a physical point of view,
  ground states with respect to $S$ are sampled,
  i.e., configurations $\myvec{y}$
  with the lowest values of $S=S(\myvec{y})$. Thus, the left tail
  of $Q(S)$ is addressed.
\item In contrast to  the physical canonical ensemble, also temperatures
  $\Theta <0$ make sense. Now, the bias is larger for configurations
  where $S=S(\myvec{y})$ is larger than typical values of $S$.
\item In particular for $\Theta \to 0^-$, configurations
  $\myvec{y}$
  with the largest values of $S$ are preferred. This addresses
  the right tail of $Q(S)$.
\end{itemize}

During the generation of
configurations $\myvec{y}$, one measures always the
corresponding values $S=S(\myvec{y})$, which finally can be collected
in histograms which approximate the distributions $Q_{\Theta}(S)$ of $S$
in the biased ensemble at temperature $\Theta$. The results will
depend of $\Theta$. This means that the histograms will cover a certain
range of values of $S$. Therefore, one usually has to
generate configurations for  several values of $\Theta$
in order to obtain a full or at least a large range
of values of $S$. Details on how the final estimate for the
distribution $Q(S)$ can be
obtained from combining the result generated at different
values of $\Theta$ is presented in Sec.~\ref{sec:glueing}.

The generation of configurations according to the biased
distribution $P_{\Theta}(\myvec{y})$ can typically not be performed
by direct sampling, i.e.\ the inversion method cannot be applied.
Thus, we use a Markov chain Monte Carlo
approach, which is very general, as we have learned
in Sec.~\ref{sec:markov}. Specifically, we use the
versatile Metropolis-Hastings algorithm as introduced in
Sec.~\ref{sec:metropolis}. Here, we assume
that the Monte Carlo trial moves, as described by the
matrix $A(\myvec{y} \to \myvec{z})$,
correspond to the original probabilities $P(\myvec{y})$.
Thus, if one  performed the trial moves and accepted them
all, the original distribution would be obtained.
This means, detailed balance holds for this part, i.e.\ 
$P(\myvec{y})A(\myvec{y}\to \myvec{z})=P(\myvec{z})A(\myvec{z}\to \myvec{y})$.

Now, not all trial configurations are accepted, instead
a Metropolis criterion is imposed that, as we now will see,
depends on the bias.
When using  Eq.~(\ref{eq:metropolis}) for
the Metropolis criterion with the target distribution $P_\Theta(\myvec{y})$,
and inserting Eqs.~(\ref{eq:biased:target:distr}) and
(\ref{eq:canonical}),  we obtain
\begin{eqnarray}
\tilde{W}(\myvec{y}\to \myvec{z}) &\stackrel{(\ref{eq:metropolis})}{=}& 
\min\left(1, \frac{P_\Theta(\myvec{z})}{P_\Theta(\myvec{y})}
\frac{A(\myvec{z}\to \myvec{y})}{A(\myvec{y}\to \myvec{z})} \right)
\nonumber \\
& \stackrel{(\ref{eq:biased:target:distr})}{=} & \min\left(1,
  \frac{P(\myvec{z}) B_{\Theta}(\myvec{z})}
       {P(\myvec{y}) B_{\Theta}(\myvec{y})}
  \frac{P(\myvec{y})}{P(\myvec{z})} \right) \nonumber \\
  & = &
  \min\left(1,\frac{B_{\Theta}(\myvec{z})}
       {B_{\Theta}(\myvec{y})} \right) \nonumber \\
       & \stackrel{(\ref{eq:canonical})}{=} &
       \min\left(1,\exp(-(S(\myvec{z})-S(\myvec{y}))/\Theta ) \right)\,
\label{eq:metropolis:bias}
\end{eqnarray}

Note that, depending on the model, it is not necessarily the
most efficient choice to generate the trial configurations 
$\myvec{z}$ such that the original distribution is reproduced.
This choice might lead to low acceptance probabilities of the
Metropolis steps. Sometimes it is better
to generate trial configurations where one has
the biased distribution already in mind. In this case, the
Metropolis acceptance probability will look different, depending on the
model and the details of the trial-configuration generation.

Nevertheless, generating the trial configurations compatible with
the original  distribution $P(\myvec{y})$ is often beneficial. For this case,
the algorithm is summarized here, $n_{\rm mc}$ being the number
of MC steps:

\begin{algorithm}{biased-MC($n_{\rm mc}$, $\Theta$)}
\> generate start configuration $\myvec{y}(0)$\\
\> {\bf for} $t = 1,\ldots, n_{\rm mc}$\\
\> {\bf do}\\
\>\> 
generate trial configuration $\myvec{z}$ from $\myvec{y}(t-1)$
compatible with $P(\myvec{z})$\\
\>\> compute $\Delta S = S(\myvec{z})-S(\myvec{y}(t-1))$\\
\>\> $r$: uniform $U(0,1)$ random number\\
\>\> {\bf if} $r<\min\left(1,\exp(-\Delta S/\Theta )
\right)$\\
\>\> {\bf then}\\
\>\>\> \myvec{y}(t)$ := \myvec{z}$\\
\>\> {\bf else}\\
\>\>\> \myvec{y}(t)$ := \myvec{y}(t-1)$\\
\> {\bf done}\\
\end{algorithm}

In general, how the trial configuration is generated from the current
configuration depends a lot on the model, and for most models
there are still many ways. Typically, only small changes of the
configuration lead to sufficiently large acceptance probabilities.
As an example, we now consider again the Bernoulli process.

\subsection{Bias for the Bernoulli Process}

For the Bernoulli process of $n$ coin flips,
the MCMC algorithm redraws in each
step a number $n_{\rm c}$
of randomly chosen elements of the current configuration $\myvec{y}(t)$,
with the original coin probability $\alpha$ as of Eq.~(\ref{eq:coin:toss}).
For the changed
elements, the current values are stored in a vector, i.e.~an
array {\tt old[]}. This allows
to restore the previous configuration, if the trial configuration
is rejected. Note that for restoring the configuration, the current
values are restored in reverse order, to prevent a mistake if
some elements are by chance selected more than once for a change.

We also assume that the quantity $S$ of interest is a general function
of the configuration $\myvec{y}$. The MCMC algorithm for
performing $n_{\rm mc}$ steps for the
Bernoulli process, performed at value $\Theta$ for the temperature-like
parameter,
reads as follows: It is assumed that the initial
or current configuration is passed as vector 
$\myvec{y}$ to the function, all necessary parameters, and the
score evaluation function $S(\cdot)$.

\begin{algorithm}{biased-MC-Bernoulli($\myvec{y}$, $n$, $n_{\rm mc}$, $n_{\rm c}$,
    $\Theta$,
    function $S(\cdot)$)}
\> {\bf for} $t = 1,\ldots, n_{\rm mc}$\\
\> {\bf do}\\
\>\> $S_{\rm old}=S(\myvec{y})$\\
\>\> {\bf for} $c=1,\ldots,n_{\rm c}$\\
\>\> {\bf do}\\
\>\>\> i = random in $\{1,\ldots, n\}$\\
\>\>\> old[$c$] = (i, $y_i$)\\
\> \>\> $r$: uniform $U(0,1)$ random number\\
\>\>\> {\bf if} $r < \alpha$ {\bf then} $y_i=1$ {\bf else} $y_i=0$\\  
\>\> {\bf done}\\
\>\> $S_{\rm new}=S(\myvec{y})$\\
\>\> compute $\Delta S = S_{\rm new}-S_{\rm old}$\\
\>\> $r$: uniform $U(0,1)$ random number\\
\>\> {\bf if} $r\ge \min\left(1,\exp(-\Delta S/\Theta )
\right)$\\
\>\> {\bf then} ``reject'' \\
\>\>\> {\bf for} $c=n_{\rm c},n_{\rm c}-1, \ldots,1$\\
\> \>\> {\bf do}\\
\> \>\>\> restore $y_i$ where $i$ and $y_i$ are stored in old[$c$]\\
\> \>\> {\bf done}\\
\> {\bf done}\\
\end{algorithm}

Note that in case of acceptance, nothing has to be done, since
the trial state is stored directly in $\myvec{y}$.
The C code \cite{bernoulli_code2024} can be found in {\tt mc\_bernoulli.c} as
function {\tt bernoulli\_mc\_step\_bias()}.

\ifthenelse{\boolean{showcode}}{
\newpage
{\small
\begin{verbatim}
int  bernoulli_mc_step_bias(int n, double *y, double alpha, int num_changes,
                          double theta, double (* calc_S)(int n, double *y))
{
  int change;                                         /* change counter */
  int pos;                                           /* where to change */
  int S_old, S_new;                    /* current/trial "energy" values */
  int *change_pos;               /* remember which entries were changed */
  double *change_val;                           /* remember old entries */
  double prob;                       /* Metropolis-Hastings probability */
  int accept;                                     /* is step accepted ? */

  change_pos = (int *) malloc(num_changes*sizeof(int));
  change_val = (double *) malloc(num_changes*sizeof(double));

  S_old = calc_S(n, y);
  for(change=0; change<num_changes; change++)        /* several changes */
  {
    pos = (int) floor(n*drand48());      /* select change position, for */
                                /* simplicity, allow for double changes */

    change_pos[change] = pos;                /* remember previous value */
    change_val[change] = y[pos];
    if(drand48() < alpha)                    /* redraw state at positon */
      y[pos] = 1;
    else
      y[pos] = 0;
  }
  S_new = calc_S(n, y);

  prob = exp(-(S_new-S_old)/theta);
  if(drand48() > prob)                                      /* reject ? */
  {
    accept = 0;
    for(change=num_changes-1; change >=0; change--)   /* reverse order! */
      y[change_pos[change]] = change_val[change];
  }
  else
    accept = 1;

  free(change_pos);
  free(change_val);
  return(accept);
}
\end{verbatim}
}
}{}

As an unusual example for the measurable quantity $S$, 
we consider the number $S_{3+}$  of blocks of at least 3 1's
in a row. For example,
the configuration
$\myvec{y}=010110$$\underline{111}$$00$$\underline{1111}$$01$ has $S_{3+}=2$.

With the \verb!mc_bernoulli.c! code \cite{bernoulli_code2024}
we can generate data for $S_{3+}$,
The Metropolis algorithm is chosen by
the \verb!-alg 2! option.
 First, we simulate  the almost unbiased case for which we choose
 $\Theta = 1000$. For this case, 
 a large number $n_{\rm c}=45$ of changes are performed
 in each step. This is
 stated with the \verb!-change! option when calling the program.
 Here, we consider $n=51$ coin flips for $\alpha=0.5$, which are
 the third and second last arguments, respectively.
 The number of Monte
 Carlo steps is the last argument, 100000 here.
 The first 10000 steps are for equilibration and therefore
 ignored for  measuring
 the histogram,
 as stated by the argument for the \verb!-histo! option.
 When calling the program in Unix from a command
 line, this can be achieved by entering
\begin{verbatim}
mc_bernoulli -S3+ -alg 2  -histo 10000 -theta 1000 \
-change 45 51 0.5 100000 > bernoulli_S3p_t1000.histo
\end{verbatim}

Note that the results are redirected to the file
{\tt  bernoulli\_S3p\_t1000.histo}.
To next access the upper tail, we choose $\Theta = -0.5$. Here only
5 elements of the coin flip vector are redrawn in each MC step,
to allow for a considerable acceptance probability.
{
\begin{verbatim}
mc_bernoulli -S3+ -alg 2  -histo 10000 -theta -0.5 \
-change 5 51 0.5 100000  > bernoulli_S3p_t-05.histo
\end{verbatim}
}
\begin{sloppypar}

\begin{figure}[ht]
\begin{center}
\includegraphics[width=0.55\textwidth]{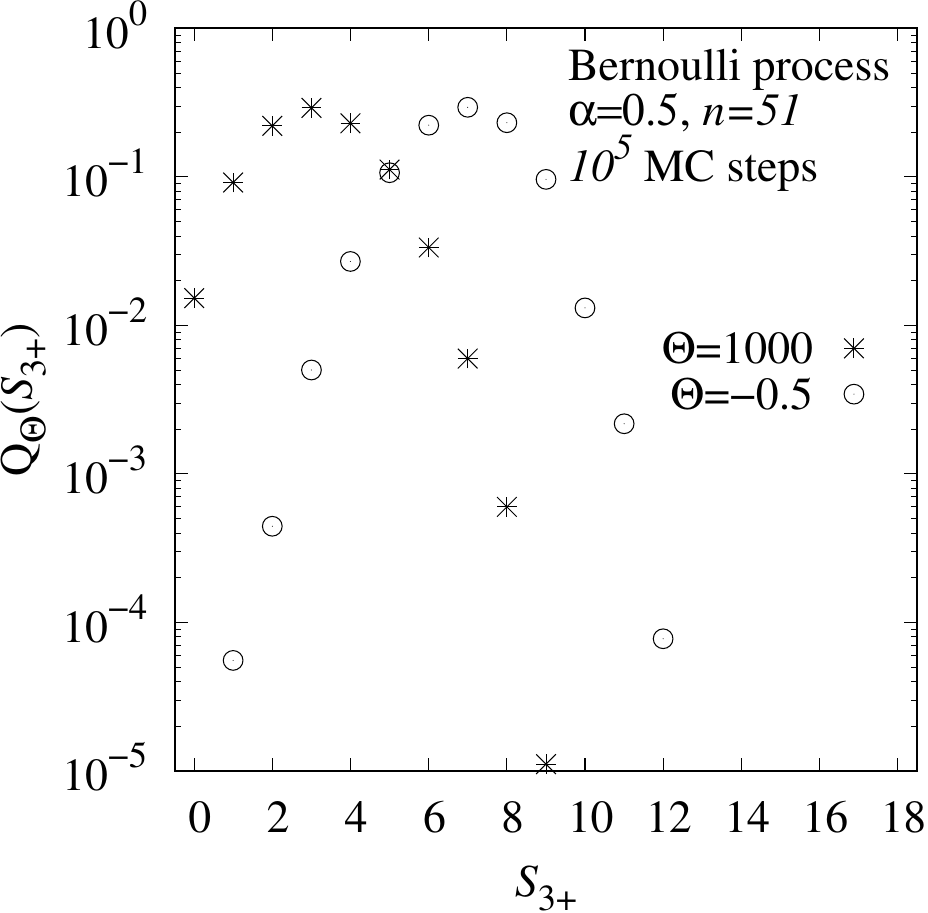}
\end{center}
\caption{Histograms of biased MCMC sampling of $S_{3+}$ for the
  Bernoulli process case ($n=51$ coin flips at $\alpha=0.5$).
  Two cases of the bias temperature $\Theta=1000$ and $\Theta=-0.5$
  were considered.
  \label{fig:mc_bernoulli_raw}
}
\end{figure}

  The two resulting histograms are shown in
Fig.~\ref{fig:mc_bernoulli_raw}. The typical-case sampling is obtained
for $\Theta=1000$, because the bias is almost 1. This
results in observed values $S_{3+}\in [0,8]$. For $\Theta=-0.5$, the
bias shifts the observed distribution to larger values 
up to $S_{3+}=12$. Note that for $n=51$ the maximum
possible value is $S_{3+}=13$. Thus, almost the full support
of the distribution $Q(S_{3+})$ is sampled.

\end{sloppypar}

You should perform simulations yourself using
the code and observe how varying the parameters, like the number
$n_{\rm c}$ of redrawn coin flips, lead to
changes of the results. In particular the empirical acceptance probability
is affected, which is
also written into the output file. It is also interesting to change $\Theta$
and try to sample the largest possible values, which is here $S_{3+}=13$.

Note that the $S_{3+}$ quantity has the property that it does not depend
mo\-no\-tonously
on the model parameter $\alpha$. If one tried, in the spirit
of the educated bias as shown in Sec.~\ref{sec:biased:sampling:bernoulli},
to simulate the process
for parameter values $\beta\neq \alpha$, one would observe:
\begin{itemize}
\item  
  If $\beta < \alpha$ is small, i.e., $\beta < \alpha$.
  there will be fewer 1's in the configuration, thus $S_{3+}$ is
  often  zero, e.g.

  $\myvec{y}=0000100010100100$ has $S_{3+}=0$.

  This could be used to access the left tail.

\item To obtain a larger number of $S_{3+}$ one
  has to use not too-small values of the parameter.
  For $\beta \approx \alpha$ there might
  be higher than typical values of $S_{3+}$. E.g., 

  $\myvec{y}=1011101011101110$ exhibits $S_{3+}=3$. But this will
  occur with the typical frequency, i.e.\ not so frequently.

\item For $\beta \gg \alpha$, one obtains again lower than
  typical values of $S_{3+}$, e.g.,
  
$\myvec{y}=1111111011111111$ has $S_{3+}=2$.
\end{itemize}

 Thus, this quantity is an example
 where the \emph{educated}  approach, i.e.\ just using a different value
 $\beta$ for the Bernoulli parameter, will not allow one to reach
 the very upper tails of the distribution
$Q(S_{3+})$. Only a \emph{blind} approach, which takes automatically
care of how the configuration space influences the measured values,
will work.

Still, one has to take some care, because the rare
configurations with untypical small values of $S_{3+}$ will
consist of sequences of $n$ coin flips with unusual small
or unusual high number of 1's. For the present example, where
the system size $n=51$ is not very large, this poses
no problem. But for much larger system sizes $n$, it will
be more difficult to obtain the correct result. Here,
a single MCMC will not grant access to 
both regions within one run, because the two regions are separated
by a region where $S_{3+}$ is typically large. Therefore,
this region acts as a barrier in configuration space.
This can be improved by performing many
runs each time staring in the region of typical
configurations.  Even better would be to use an approach
like \emph{Parallel Tempering}
\cite{geyer1991,hukushima1996}. It is based on performing the
simulations for several
temperature parameters $\Theta_1,\ldots,\Theta_K$ in parallel,
while allowing the configurations to switch temperatures in
a controlled way, also governed by a Metropolis criterion.
This approach is suited to explore broad ranges of the configurations
space.

As final example, we show a measurable quantity,
where it is even clearer that the \emph{educated} approach will not work.
We consider
the number $S_{0101}$ of non-overlapping segments 0101 in $\myvec{y}$.
The following two example sequences have exactly half of the entries
1, and the other half 0:
\begin{itemize}
\item $\myvec{y}=0000000011111111$ has $S_{0101}=0$,

\item $\myvec{y}=0101010101010101$ has $S_{0101}=4$,
\end{itemize}
while the resulting values for $S_{0101}$ are very different.
Here it is even more obvious that the \emph{educated} approach
of  performing simulations
for other values $\beta\neq \alpha$ of the model parameter will
not give access to the tails of the distribution. Here, only the
more general \emph{blind} approach,
which depends on the value $S$ of interest, is suitable.

\subsection{Obtaining the true Distribution }
\label{sec:glueing}

The final step is to obtain for the quantity $S$ of interest the
estimate of the original distribution $Q(S)$, see Eq.~(\ref{eq:def:Q:S}).
Since the sum over the exponential many configurations cannot be performed,
$Q(S)$ shall be erstimated from the simulations.
The starting point for the analysis is the collection
of histograms estimating the distributions $Q_{\Theta}(S)$ in the
biased ensemble,
for different values of the temperature-like parameter $\Theta$. One has

\begin{eqnarray}
Q_{\Theta}(S) & = &
\sum_{\myvec{y}} \delta_{S,S(\myvec{y})} P_{\Theta}({\myvec{y}}) \nonumber \\
& = &
\sum_{\myvec{y}} \delta_{S,S(\myvec{y})}  \frac{1}{Z(\theta)}
P(\myvec{y}) B_{\Theta}(\myvec{y}) \nonumber \\
& = &
\frac{1}{Z(\Theta)} \sum_{\myvec{y}} \delta_{S,S(\myvec{y})}  
P(\myvec{y}) \exp(-{S}(\myvec{y})/\Theta)  \nonumber \\
& \stackrel{(\star)}{=} & \frac{1}{Z(\Theta)} \exp(-{S}/\Theta)
\sum_{\myvec{y}} \delta_{S,S(\myvec{y})}  
P(\myvec{y})   \nonumber \\
& = & \frac{1}{Z(\Theta)} \exp(-{S}/\Theta) Q(S) \nonumber \\
\Rightarrow Q(S) & = & Z(\Theta)\exp(S/\Theta) Q_{\Theta}(S)\,.
\label{eq:Q_S:unbiased}
\end{eqnarray}
For equality $(\star)$, due 
to the delta factor $\delta_{S,S(\myvec{y})}$, we
have $\exp(-{S}(\myvec{y})/\Theta)$ is equal to $\exp(-{S}/\Theta)$.
Therefore, it can be moved in front of the sum.
The results means that the true distribution $Q(S)$ can be obtained from
the biased distribution $Q_{\Theta}(S)$!
Note that the right side depends on $\Theta$, while the left does not.
Thus, theoretically, the data obtained for any value of $\Theta$
is sufficient to obtain $Q(S)$. In practice, the accumulated data
for any value of $\Theta$
will concentrate in some interval $I^{\Theta}=[S_{\min}^{\Theta},S_{\max}^{\Theta}]$,
thus $Q(S)$ can be estimated
from $Q_{\Theta}(S)$ only in this interval.
For practical reasons, we assume that this interval consist only
of values where the statistics is ``good enough'', i.e.\ does not contain
rare outliers with respect to $Q_{\Theta}(S)$. Since such an interval
is finite, to cover a large range of the support of $Q(S)$,
 one has
to perform simulations for several values of $\Theta$.

Also it is interesting that one has to divide in Eq.~(\ref{eq:Q_S:unbiased})
by $\exp(-S/\Theta)/Z(\Theta)$ which corresponds to
$B_{\Theta}(\myvec{y})$. This  is similar to
unbiasing in Eq.~(\ref{eq:ph-averageB}).
The only difference is that previously it was assumed that the
bias is fully known, while here the values $Z(\Theta)$ are not known a priori.
Next, we will discuss how these normalization constants can be obtained.

The basic idea goes as follows:
Assume that for two values $\Theta_1$, $\Theta_2$ the
intervals $I^{\Theta_1}$ and $I^{\Theta_2}$ of sufficiently sampled data  overlap,
i.e.\ $I=I^{\Theta_1}\cap I^{\Theta_2}$ is not empty. Thus, for $S\in I$,
we have substantial data to estimate both
 $Q_{\Theta_1}(S)$ and $Q_{\Theta_2}(S)$. Here one has

\begin{equation}
Z(\Theta_1)\exp(S/\Theta_1) Q_{\Theta_1}(S) =
Q(S) =  Z(\Theta_2)\exp(S/\Theta_2) Q_{\Theta_2}(S)
\label{eq:normalization:condition}
\end{equation}

Thus, the ratio of the normalization constants is fixed to 

\begin{equation}
\frac{Z(\Theta_1)}{Z(\Theta_2)}  =
  \frac{\exp(S/\Theta_2) Q_{\Theta_2}(S)}{\exp(S/\Theta_1) Q_{\Theta_1}(S)}\,.
\label{eq:normalization:conditionB}
\end{equation}

Since the sampled and normalized histograms only approximate
the distributions,
this relation will also be fulfilled only approximately. Note that for
$K$ different values $\Theta_1,\ldots,\Theta_K$ one can determine
$K-1$ ratios. In addition, the final estimate for $Q(S)$ should be normalized 
as well. This can be approximated by $Z(\infty)=1$ and
provides the final relation to fix all
normalization constants.

In order to actually arrive with the given histograms
at the estimate for the final distribution $Q(S)$, we first discuss a
manual approach. We restrict ourselves to two temperatures $\Theta_1$
and $\Theta_2$, the
generalization to more temperatures is straightforward.
This approach means, we choose $Z(\Theta_1)$ and $Z(\Theta_2)$ such that
when plotting the two rescaled histograms
they almost agree, within statistical fluctuations,
in the interval $I$ of values of $S$ where both
histograms have gathered sufficiently data.
One can say, the distributions are ``glued''
or ``stiched'' together.

\begin{figure}[ht]
\begin{center}
\includegraphics[width=0.55\textwidth]{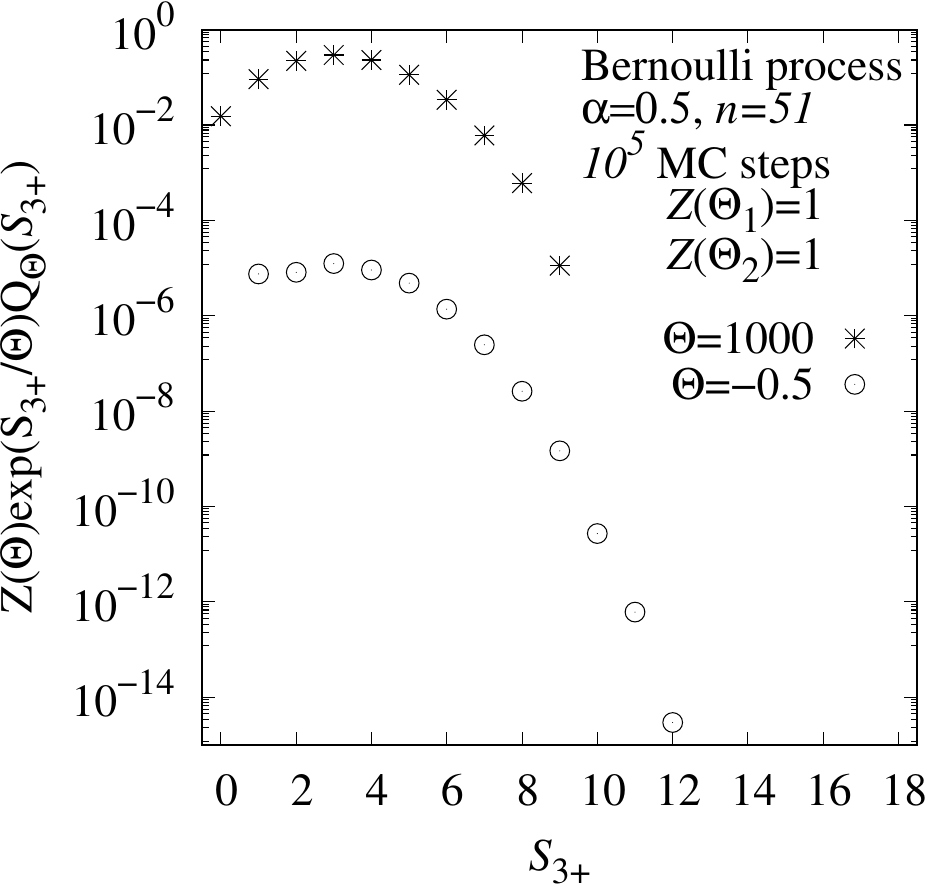}
\end{center}
\caption{Rescaled histograms of biased MCMC sampling of $S_{3+}$ for
  the Bernoulli process ($n=51$ coin flips at $\alpha=0.5$).
  Two cases of the bias temperature $\Theta=1000$ and $\Theta=-0.5$
  were considered. The initial values for the partition functions
  are $Z(1000)=1$ and $Z(-0.5)=1$.
  \label{fig:mc_bernoulli_scaled}
}
\end{figure}

We show how this works for the data as generated in the last
section, using the \verb!gnuplot! program. For the
data obtained at $\Theta_1=1000$, which is effectively infinity,
we assume  $Z(\Theta_1) \approx 1$. To start the manual
adjustment of $Z(\Theta_2)$, we also start with $Z(\Theta_2)=1$
and plot the histograms rescaled by the exponential bias with
the following \verb!gnuplot command!:

\begin{verbatim}
gnuplot> Z1=1
gnuplot> Z2=1
gnuplot> plot [-0.5:18] \
   "bernoulli_S3p_t1000.histo", using 1:(Z1*exp($1/1000)*$2) \
   "bernoulli_S3p_t-05.histo" using 1:(Z2*exp($1/-0.5)*$2)
\end{verbatim}

Note that we have used the \verb!using! data modifier, which can
be abbreviated by the letter \verb!u!. The data modifier
\verb!1:(Z1*exp($1/1000)*$2)! of the
first plotted curve, provided by the file \verb!"bernoulli_S3p_t1000.histo"!,
states
that the $x$ coordinate is just taken from the first column, while
the $y$ coordinate is the product of the value for \verb!Z1!,
times the exponential of the value in the first column (\verb!$1!)
devided by the temperature 1000, times the value in the second column
(\verb!$2!). The result will look like Fig.~\ref{fig:mc_bernoulli_scaled}.

As one can see, the shape of the rescaled histogram for $\Theta=-0.5$
looks similar to the histogram for $\Theta=1000$, but the frequencies
for $\Theta=-0.5$ are some orders
of magnitude too small. This means $Z(-0.5)$ must be larger, about
a factor of $10^5$. Now one can iteratively refine $Z(-0.5)$
and replot, until the two rescaled histograms agree well in the
overlapping region. Here one could be satisfied with 
$Z(-0.5)=2.3\times 10^{-3}$, i.e.

\begin{verbatim}
gnuplot> Z2=2.3e4
gnuplot> plot [-0.5:18] \
   "bernoulli_S3p_t1000.histo", u 1:(Z1*exp($1/1000)*$2) \
   "bernoulli_S3p_t-05.histo" u 1:(Z2*exp($1/-0.5)*$2)
\end{verbatim}

\begin{figure}[ht]
\begin{center}
\includegraphics[width=0.55\textwidth]{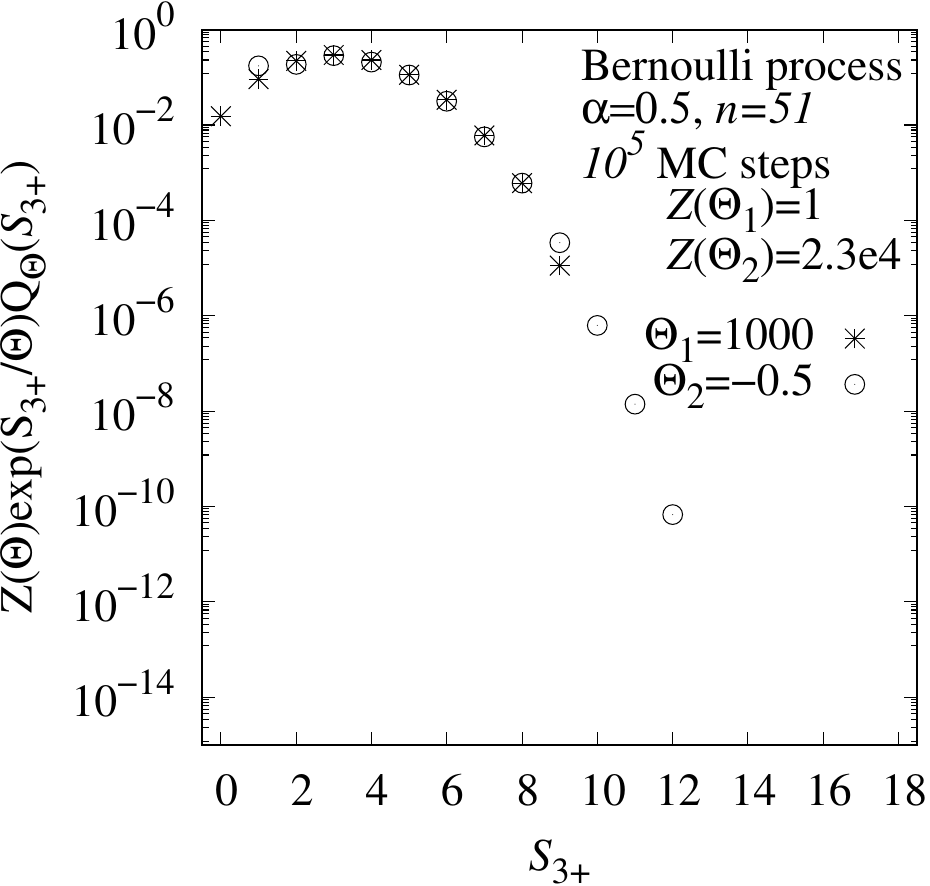}
\end{center}
\caption{Rescaled and glued histograms of biased MCMC sampling of $S_{3+}$ for
  Bernoulli process case ($n=51$ coin flips at $\alpha=0.5$).
  Two cases of the bias temperature $\Theta_1=1000$ and $\Theta_2=-0.5$
  were considered. The final values for the partition functions
  are $Z(1000)=1$ and $Z(-0.5)=2.3\times 10^{-4}$.
  \label{fig:mc_bernoulli_final}
}
\end{figure}

The result will look like Fig.~\ref{fig:mc_bernoulli_final}.
Note that the histogram for $\Theta=1000$ is actually not much rescaled
since still $Z(1000)=1$ and $\exp(-S_{3+}/1000)$ is also about 1.
Now the two rescaled histograms agree well in the middle of
the overlapping region, say $S\in[3,8]$. One could assemble
the final histogram, which is not shown here, by taking each data point from the
biased and rescaled histogram where the statistics is best. For
the present example this would be
for $S\in[0,5]$ from the $\Theta=1000$
result  and for $S\in[6,12]$ from the $\Theta=-0.5$ result.

Note that for a larger system size $n$, one needs a larger number of
temperatures. In this case one could, after gluing the results for
$\Theta=1000$ and $\Theta=-0.5$, rescale the third histogram, say obtained
at $\Theta=-0.2$ for even larger values of $S_{3+}$, such that it
matches the so far estimated distribution. For very large
systems of other models, one might have to perform simulations
for a number $K$ of temperatures up to several hundreds
\cite{kpz2018,kpz_long2020}. Here,
the calculation of all partition functions and obtaining the final
distribution by hand would be tedious. Thus, an automated approach is better
suited. This is also the case for few temperatures, because
of the higher precision.

One way to numerically rescale two histograms
obtained at neighboring temperatures $\Theta_k,\Theta_{k+1}$ for
$k=1\,\ldots,K-1$ 
such that they mostly agree, 
is to 
minimize the mean-squared difference
between the rescaled
histograms 
in the overlapping region $S \in I^{k,k+1}$. Here
we assume that $Z(\Theta_k)$ is already known; it could be
$Z(\infty)\approx 1$, or obtained from rescaling of other parts
of the distribution. 
When writing $r_{\Theta}(S) = \exp(S/\Theta) \tilde Q_{\Theta}(S)$,
where $\tilde Q_{\Theta}(S)$ is the actually obtained normalized histogram
at temperature $\Theta$,
the minimization with respect to $Z(\Theta_{k+1})$ translates into,

\begin{eqnarray}
\Delta & \equiv & \sum_{S\in I^{k,k+1}} \left(
Z(\Theta_k)r_{\Theta_k}(S) -
Z(\Theta_{k+1})r_{\Theta_{k+1}}(S) 
\right)^2 \; \to \; \min \nonumber \\
\Rightarrow 0 = \frac{d \Delta}{d Z(\Theta_{k+1})}  & = &
\sum_{S\in I^{k,k+1}} 2r_{\Theta_{k+1}}(S)\left(
Z(\Theta_{k})r_{\Theta_k}(S) -
Z(\Theta_{k+1})r_{\Theta_{k+1}}(S) 
\right) \nonumber \\
\rightarrow
Z(\Theta_{k+1}) & = & \frac{Z(\Theta_{k}) \sum_{S\in I^{k,k+1}}
r_{\Theta_k}(S)r_{\Theta_{k+1}}(S)}
{\sum_{S\in I^{k,k+1}}r_{\Theta_{k+1}}(S)^2
} \label{eq:glueing:msd}
\end{eqnarray}

It is assumed that the histograms are binned, therefore the sums
$\sum_{S\in I^{k,k+1}}$ run over the corresponding bins in the overlapping
region. As above,
 one would choose the intervals $I^{k,k+1}$  
such that they contain enough data, for both histograms.

One could start with a histogram obtained by unbiased
sampling and assume $Z(\infty)=1$ there. Then one can obtain the partition
functions
for neighboring values of $\Theta$, i.e.\ where $|\Theta|$ is rather large,
and so on.

In the version above,
all contributions to the sums are equally weighted. One could
also use, in different ways,  weights which represent the
combined qualities of the statistics for the different bins $S$.
A high quality means that both histograms for $\Theta_k$ and $\Theta_{k+1}$
contain many entries. Correspondingly, weights could be used to assemble
the final result $Q(S)$ from all histograms, such that for each value
of $S$ two or possibly even more histograms contribute. For a simple
approach, which is usually sufficient, one takes only the histograms that
have the best statistics for a given bin.

For completeness, one can evaluate the normalization factor, i.e., the
integral, of the final result and divide by it to get a final normalization
of exactly one.

There are alternatives for obtaining the normalization constants.
One could, e.g.\ calculate relative normalization constants
$Z(\Theta_k)/Z(\Theta_{k+1})$
directly using Eq.~(\ref{eq:normalization:condition}) for 
single bin values
$S \in I^{k,k+1}$ directly. This gives a certain number of estimates for the
ratio of the normalization constants. The final result
for $\Theta_k$ and $\Theta_{k+1}$
 is then obtained
by averaging over these estimates, possibly weighted by the
accuracy of the estimates, i.e., by the statistics of the contributing
bins.

Even better would be to apply the 
multi-histogram reweighting approach from Ferrenberg and Swendsen
\cite{ferrenberg1989}. It  includes precision/statistics of
different histogram entries and therefore allows for calculation of
error bars. Note that you do not have to implement the approach
yourself, because there is a Python-based tool by
Peter Werner \cite{werner2022} which does the job for you.


\subsection{Generalization and other examples}

For the so-far used example, the Bernoulli process with parameters
probability $\alpha$
and size $n$, the state
of the Markov chain was given by a vector $\myvec{y} \in \{0,1\}^n$.
From the given vector at MC time $t$ the quantity of interest $S(t)=S(\myvec{y}(t))$
was calculated. Along the Markov chain
$\myvec{y}(0) \to \myvec{y}(1) \to \ldots$ typically
some entries of the state changed from time step $t$ to time $t+1$,
with corresponding changes of the quantity $S$.
This can be depicted as follows:

\begin{equation*}
\begin{array}{cccc}
  \fbox{\,\,\myvec{$y$}(t)\,\,} & \longrightarrow  & \fbox{ \myvec{$y$}(t+1)}
  & \longrightarrow\\
  \Downarrow & & \Downarrow \\
S(t) && S(t+1) 
\end{array}
\end{equation*}
where the $\longrightarrow$ symbol indicates the random evolution
of the states while the $\Downarrow$ symbol represents
a deterministic evaluation.

Within the Metropolis algorithm a trial
state $\myvec{z}$ is obtained by first copying the
current state $\myvec{y}(t)$ and then changing some randomly chosen
entries $z_i$. The new entries are drawn with the original statistics,
i.e., each time a uniformly $U(0,1)$ distributed random number $\xi$ is
drawn, and then $z_i=1$ if $r<\alpha$ and $z_i=0$ otherwise.

Note, it would be completely equivalent, if instead of storing the result
$y_i$ of a coin flip, we store the random numbers used to generate
the coin flip in a vector
$\myvec{\xi} \in [0,1]^n$, i.e.\ $\xi_i$ is the random number for the
$i$'th coin flip. Then one could
deterministically assign the coin-flip vector
$\myvec{y}=\myvec{y}(\myvec{\xi)}$
by $y_i=1$ if $\xi_i<\alpha$ and $y_i=0$ else. Thus, instead
of evolving a Markov chain of the coin flip vectors $\myvec{y}(t)$,
from which $S=S(\myvec{y}(t))$
one could evolve a Markov chain of random vectors $\myvec{\xi}(t)$
and obtain $S=S(\myvec{y}(\myvec{\xi}(t))$, i.e. $S=S(\myvec{\xi}(t))$,
with formally a different function dependency of $S$
on $\myvec{\xi}$ as compared to $S$ on $\myvec{y}$,
here denoted for simplicity both by $S(\cdot)$.
The modified approach can be depicted
as follows:

\begin{equation*}
\begin{array}{cccc}
  \fbox{ \,\,\myvec{$\xi$}(t)\,\,} & \longrightarrow  &\fbox{ \myvec{$\xi$}(t+1)}
  & \longrightarrow\\
  \Downarrow & & \Downarrow \\
\myvec{y}(t) && \myvec{y}(t+1) \\
  \Downarrow & & \Downarrow \\
S(t) && S(t+1) 
\end{array}
\end{equation*}

Clearly, both approaches, having a Markov chain of states $\myvec{y}(t)$,
and having a Markov chain of states $\myvec{\xi}(t)$, are 
equivalent, only that the latter one looks more complicated. Nevertheless,
the latter one is more general! The reason is that whatever
stochastic model you investigate, and whatever random quantities you calculate
within the model, the implementation will always be based
on $U(0,1)$ uniformly distributed random numbers. Some of these numbers
may, depending on the model,
be turned by the inversion method to exponentially distributed
numbers, some by the Box-Muller approach to Gaussian numbers,
some random numbers
may be used to decide the random orientation of some spins,
or the random positions of nodes in a plane, and so on. This means,
basically the implementation of any random process can be represented
as a, often complex, deterministic
mapping from a vector of $n$ uniformly distributed
random numbers to a state of the system. Then, from the state some
quantity $S$ of interest can be calculated,
in a more or less complex way. Thus, for any random
process, the quantity of interest is a deterministic function
$S=S(\myvec{\xi}(t))$ of a vector $\myvec{\xi}$ of random numbers.
On the implementation side, this means one disentangles the
generation of the random numbers from the other ingredients of a model.
Normally, one would call a random number generator at different
places in the code. With the disentangles approach, one
would first generate all $U(0,1)$ distributed (pseudo) random
numbers, store them in the vector and use them when needed,
while the other parts of the simulation are performed.

In this way, one can still easily implement direct sampling,
by first randomly choosing
all entries of $\myvec{\xi}$ uniformly in $[0,1]$ and then
evaluating $S$. The histogram of observed
values of $S$ will give an estimate of the underlying distribution $P(S)$
in the high-probability region.
On the other hand, if one is interested in the tails of $P(S)$
one can introduce a bias, say proportional to $\exp(-S(\myvec{\xi})/\Theta)$
with parameter $\Theta$, and then use a Markov chain Monte Carlo simulation
to drive the simulation to the rare values of $S$, as it has been demonstrated
so far. This means, the large-deviation approach presented
here is very general
and can be applied, at least in principle, to any stochastic system,
equilibrium and non-equilibrium ones, from various fields of science.

Consequently, the large-deviation behavior of a large variety of models
has been studied with this approach.

\subsubsection{Random walk}

A simple example is the random walk, which is just the
sum of, typically independent, random numbers $y_i$ called \emph{increments}.
The increments can have arbitrary distribution, e.g., discrete,
Gaussian, exponential or power-law. Thus, one can generate
a $n$-step random walk from a given
vector $\myvec {\xi}$ of uniformly $U(0,1)$
distributed random numbers by generating random numbers $y_i$ from
the desired distribution and then summing. The resulting position at time
$t\le n$ is given by $x(t)=\sum_{i=1}^t y_i$.
Standard random walks
are well understood, i.e. $P(x)$ is well known.
More interesting are extensions, e.g., when the increments are
correlated. One special variant, with a power-law correlation,
is called \emph{fractional Brownian motion} \cite{mandelbrot1968}.
Here, depending
on the so called \emph{Hurst} exponent $H$, the increments are positively
($H>1/2$), negatively $(H<1/2)$ or not ($H=1/2$) correlated.
The correlation can be conveniently imposed \cite{wood1994,dietrich1997}
by using a 
Fourier Transform (FT) of the correlation function plus a back transformation
of the product of the randomness and the FT correlation.
This process becomes even more challenging
when one
introduces an absorbing boundary at, say, $x=0$. This means that
only walks are considered where for all times $t$ the position $x(t)$
is not negative. Here, the distribution $P(x)$ for $x=x(n)$  is of interest
near $x=0$, since here the probabilities are very small, due to
the absorbing boundary. Unfortunately, the direct sampling approach
will generate many walks where at some time $x(t)<0$ and thus one
has to disregard the walk. The probability of not being disregarded,
the so-called \emph{persistence}, decreases like a power law
$\sim t^{H-1}$. Here, a MCMC approach comes handy even
without a bias, because one can restrict the Markov chain to walks
which do not get absorbed. This can be achieved within
the Metropolis algorithm by rejecting any trial
configuration, where the resulting walk $x(t)$ shows an absorption.
By using in addition also a bias $\sim \exp(-x(n)/\Theta)$,
the low probability tail of $P(x)$ has been determined successfully
within a large-deviation study \cite{fBm_MC2013}. This allowed
the authors to numerically verify that indeed $P(x)$ follows
a power law $\sim x^{\phi}$ near $x=0$,
where the power $\phi$ depends on the Hurst exponent like $\phi=(1-H)/H$
as predicted analytically \cite{zoia2009, wiese2011}.

\subsubsection{Non-interacting Fermions}

Also simple by its definition, but slightly more involved
with respect to the large-deviation approach, is the following model.
It describes the ground state of $K$ non-interacting fermions in a
random energy 
landscape \cite{derrida1981} of
$n$ sites. The potential landscape is just a vector $\myvec{y}$
of $n$ independently and identically distributed
random numbers $y_i$
following an arbitrary given distribution, e.g.\ exponential.
Thus, a realization of the landscape can  again
be generated easily from a vector
$\myvec{\xi}$ of $n$ uniform $U(0,1)$ random numbers.  
The fermions occupy the $K$ lowest energy levels, thus the ground
state energy $E_0$ is obtained by sorting the energies in $\myvec{y}$
and summing up the lowest $K$ values. The tails of the
distribution $P(E_0)$ have also been addressed \cite{srn2018}
by a large-deviation
approach with a bias $\sim \exp(-E_0/\Theta)$. Here, the
difficulty arises that $E_0$ depends a lot on the actual entries
in $\myvec{\xi}$. The far tails are governed by vectors $\myvec{\xi}$ where
at least $K$ entries are all very close to $0$ or
all very close to $1$. Here, redrawing entries $\xi_i$ completely
in $[0,1]$
would change the resulting value $E_0$ too much, leading to
too many rejections within the Metropolis algorithm, i.e.
no progress. Instead an approach was used, where for the generation
of the trial states one starts by selecting a random
magnitude $\delta$ among six values
$\delta \in \{10^0,10^{-1},\ldots, 10^{-5}\}$. Then
some randomly chosen entries $\xi_i$ are changed only
a bit by applying $\xi_i=\xi_i+\epsilon \delta$,
where $\epsilon$ is uniformly  $U(-1,1)$ distributed.
 Moves leading
 $\xi_i$ outside the interval $[0,1]$ lead also to an immediate rejection
of the change of the entry $\xi$.
Thus, the entries $\xi_i$ perform random walks within the
interval $[0,1]$ with reflecting boundary conditions,
which guarantees a uniform distribution in $[0,1]$. If the
actual composition of $\myvec{\xi}$ is crucial, the moves
where $\delta$ is too large lead to too large changes of $E_0$
and are therefore rejected. But there will be always trial-state
generations
where $\delta=10^{-4}$ or $\delta=10^{-5}$. These are typically
accepted and lead to a slow but steady evolution of the states
within the Markov chain. By using this approach, the distribution
$P(E_0)$ of the ground states energies could be determined
down to probabilities as small as $10^{-160}$.

\subsubsection{Traffic model}

A more complex model is, e.g.\ the Nagel-Schreckenberg model \cite{nagel1992}.
It describes a one-lane
one-directional traffic of cars, where the density $\rho$ of cars is the
control parameter. The cars accelerate in each time step a bit
if possible or
break if they come too close to cars in front of them. In addition
to these deterministic rules, there is some  random breaking,
even if there is enough space. For these random breaking events,
collected along a full
temporal development, a
vector $\myvec{\xi}$ of random number is used, which defines
the complete randomness of the dynamics. The distribution
$P(q)$ of the traffic flow, measured at the end of the dynamical evolution
over some time, was investigated. By considering an exponential
bias $\sim \exp(-q/\Theta)$, performing a Markov chain in which the
randomness $\myvec{\xi}$ represents the state, and
subsequent unbiasing $\sim \exp(+q/\Theta)$,
the distribution $P(q)$ could be determined
\cite{nagel_schreckenberg2019}
down to probability densities such as $10^{-160}$. In particular it
was observed that the shape of the tails of $P(q)$ change, depending
on the density $\rho$, whether the system is in the low-density
high-flow phase, or in the medium or high-density congested phase.

\subsubsection{Spread of diseases}

Also many models of the evolution of the spread of a diseases on
networks feature stochastic dynamics. The nodes
of the network represent individuals and the edges personal contacts. For
the well known \emph{susceptible-infected-recovered} (SIR) model
\cite{kermack1927},
each node can be in one of these three states. An infected node
transfers the disease with some rate or probability $\gamma$ to
neighboring nodes that are susceptible. In addition, infected nodes
can recover with another rate or probability $\beta$. Typically,
the dynamics starts with one or few infected nodes, and stops when
no infected nodes remain. The randomness, to decide whether
nodes switch state, i.e. all random ingredients of a fully dynamical
evolution, can also be stored in a vector $\myvec{\xi}$.
This allows for a biased Marcov chain simulation. 
Using this approach, large-deviation studies \cite{sir_rare2022} have been
performed in particular to investigate the distributions $P(I)$
of the number $I$ of nodes that caught the infection for
some time. Not only the distribution $P(I)$ itself was
of interest, but also what type of dynamical evolution lead
to particular rare but severe outbreaks. Also the influence
of vaccinations \cite{sir_vaccine2023}, the effect of lock downs
\cite{sir_lockdown2023}, or how
wearing masks \cite{sir_masks2023} may limit the spread of the disease
have been studied within large-deviation frameworks.

\subsubsection{Stochastic Thermodynamics}

Also of interest are processes involving the generation of
physical work. They have been increasingly studied in the context
of \emph{stochastic thermodynamics}. One setup of interest is where
one starts a system coupled to a heat bath held at a physical
temperature $T$
in equilibrium. Then one changes, quickly or slowly,
an external parameter $B:B_0 \to B_1$
while the system is still in contact with the heat bath. Changing $B$
leads to some work $W$ being generated. When $B=B_1$ the process
ends in a non-equilibrium
state of the system. Interestingly, by the equations
of Jarzynski \cite{jarzynski1997} and Crooks \cite{crooks1999} one
is able to obtain from the distribution $P(W)$ of work
the free energy difference $\Delta F$
between the states at $B=B_0$
and $B=B_1$. Note that $\Delta F$ is an equilibrium quantity, while
the process and $P(W)$ are non-equilibrium.
 Importantly, in particular if the system consist of more
than few particles, one needs to know $P(W)$ down to the very
low probability tails in order to determine $\Delta F$.
Consequently, for numerical simulations
a large-deviation algorithm should  be used. These processes
typically have several random contributions: the sampling of the
initial equilibrium
state and the thermal noise which appears while the non-equilibrium process is
performed. Thus, the mapping from the vector $\myvec{\xi}$ to
the final result $W$ may be quite involved and may require a substential
amount of simulation time. This time has to be spent
for each single step of the Markov chain which evolves the
randomness vector $\myvec{\xi}$. The approach has been applied
to non-equilibrium work processes of the Ising model in an external
field \cite{work_ising2014} and the unfolding of RNAs by an external force
\cite{work_rna2021}. In both cases $P(W)$ could be obtained down
to sufficiently
small probability densities like $10^{-100}$ and the equations
of Jarzynski \cite{jarzynski1997} and Crooks \cite{crooks1999}
could be confirmed.

\subsubsection{Direct encoding of configurations}

Thus, using a vector $\myvec{\xi}$ of random numbers
to separate the randomness from the deterministic evolution
of a system is very convenient. Nevertheless, there are still many examples
where the large-deviation properties of a system have been
studied by performing a Markov chain with the physical systems
being directly used as the state of the chain. Examples are the
\begin{itemize}
\item distribution of sequence-alignment scores
    \cite{align2002,align_long2007}, 
\item  properties of random graphs
\cite{largest-2011,distr_qcore2017,diameter2018},
\item stability of steady-state
energy grids \cite{resilience2014,power_flow2015},
\item distribution of the length and entropy of longest increasing
  subsequences for
  permutations of random numbers \cite{lis2019,lis_count2020},
\item the distribution of the free energy of 
directed polymers in disordered media
  \cite{kpz2018,kpz_long2020,kpz_shape2021}
  \item ground state energy of random magnets \cite{pe_sk2006}.
\end{itemize}

Many more systems can be studied with respect to large-deviations
properties. Let us hope that this introduction has equipped you
with the foundations allowing you to enter the field sucessfully!

\subsection{Computational resources}

One might wonder what computational resources are needed to study
the tails of the distribution. First, most of the time is spent
in running single instances of the model of interest, for a given
set of random numbers, or for a given directly encoded instance. This sets
the time scale needed to perform the MCMC simulations, say for
an exponential bias $\exp(-S/\Theta)$ with a given temperature $\Theta$.
A typical number of MC steps is $O(10^6)$. Often a run for a single instance
requires only few milliseconds, if order of 1000 variables
describe the instance. This means the biased MCMC will run few 1000s,
which can be
done on a laptop, as it is the case for
the simple Bernoulli process considered here.
If one single runs requires
one second, the MCMC will take correspondingly more.

Second, one has to take into account how many of temperature values
$\Theta$ have to be
considered, since the simulations have to be performed independently for all
of them. Typically, to reach probabilities as small as $10^{-50}$,
of the order of 10 temperatures are needed. If one wants to go much
lower, many more temperatures are needed. For example in the study
of the directed polymers in disordered media
\cite{kpz2018,kpz_long2020,kpz_shape2021}, probabilities as small
as $10^{-1000}$ were obtained, which required more than 200
different  values of $\Theta$. Since also the system sizes
were rather large, leading to longer running times to evaluate
single instances, a parallel implementation on a high-performance cluster
was used and required
several weeks while  running on few hundreds of cores in parallel.

\subsection{Other approaches}

The presented approach of changing random numbers and using an exponential
bias is rather general but not completely universal, so other approaches
exist.

Sometimes, in particular if $P(S)$ is not concave,
the exponential bias $\exp(-S/\Theta)$
leads to an effective distribution $P(S)\exp(-S/\Theta)/Z(\Theta)$ which
is not bell-like around a typical value as shown
in Fig.~\ref{fig:mc_bernoulli_raw}, but instead could exhibit
a two-peak structure \cite{largest-2011}. In this case the values of $S$
between the two peaks are hard to access. This might turn
even impossible if the
system size is large. One way out could be to use a more concentrated bias,
which forces the biased distribution to be bell-like. One could, e.g.,
use a Gaussian bias \cite{neuhaus2006} $\exp(-(S-S_0)^2/(2\sigma^2))$,
as it has recently been applied to study the distribution of susceptibilities
for diluted magnetic systems \cite{griffiths_rbim2024}.

An even more general approach is to aim at using $1/P(S)$ as the bias.
Since the resulting distribution is proprtional to $P(S)/P(S)=1$, i.e.,
one obtains a more or less uniform sampling. This approach is called
\emph{Umbrella sampling} \cite{torrie1977}. Nevertheless, typically $P(S)$ is now
known. In this case one tries to construct $P(S)$ iteratively with
increasing accurcy during the
simulation, starting from a flat distribution. For this purposes
methods like the
\emph{Wang-Landau} approach are useful
\cite{wang2001}, which was applied to study rare trajectories of infection
dynamics \cite{sir_rare2022,sir_vaccine2023,sir_lockdown2023,sir_masks2023},
as already mentioned above. A related
algorithm is the
\emph{Multi-canonical Ensemble}
\cite{berg1992} which has been applied for the rare-event case
to testing error-correcting codes
\cite{iba2008}. 
Another issue could be that the system of interest is simulated
with a given package, which is hard or even impossible to change,
such that one cannot replace the internal calls to a random number generator
by accesses to a vector of random numbers. Also, sometimes the
actual dynamics is free of randomness, i.e., deterministic.
In both cases, one could
perform small random changes to the initial  or
intermediate states of the system
under consideration. Such approaches are used for rare-event sampling
of climate models \cite{lePriol2024}. Here, the single runs of the original
model are so demanding that indeed a MCMC sampling with each step
being a full run of the climate model over the full period of interest
would require too much time. Here, in particular in the case
where the quantity $S$ of interest is integrated over time., e.g.,
the physical temperature in the climate model,
population based approaches, so called \emph{Cloning Algorithms} 
\cite{giardina2011},
sometimes referred as genetic selection algorithms,
are convenient.
Here, instances with slightly different initial conditions
are simulated, and after a suitable number of steps for the model,
the population is reshaped according the biased probabilities
of the different population members. Such approaches have also
been used for some models, like the \emph{Totally Asymmetric
Exclusion Process} (TASEP) \cite{giardina2006}.

For other problems, sometime one is not interested in particular rare
endpoints of a process, but in rare trajectories between known
endpoints. This is the case when considering configurational changes
of proteins. Here the \emph{Transition-Path Sampling Approach}
\cite{dellago1998,bolhuis2002} is very useful.

\section{Conclusion}
This text provides the fundamentals to access the tails of distributions
in simulations of stochastic systems. Then main idea of the
central approach is to separate
the randomness from the othergredients of the model and keep the randomness
in a vector of $U(0,1)$ distributed entries.
This allows one to
perform a Markov Chain Monte Carlo simulation including a bias,
which drives the simulation to the desired range of values. Particularily
useful is the exponential bias, with a temperature-like parameter
controlling the range of values. By performing simulation for different
temperatures, often the distribution can be obtained numerically
over hundreds of decades in probability. Here, the methods are illustrated
using a very simple model, the Bernoulli process. Since any stochastic
simulation relies in principle on $U(0,1)$ random numbers, the approach
is very general and can be applied to a wide range of
equilibrium and non-equilibrium models. Still, many issues can arise,
like choosing a suitable bias or to determine efficient moves when
changing the configurations in the Markov chain.

\section{Acknowledgements}
The authors thanks Abishek Dhar, Joachim Krug, Satya N. Majumdar, Alberto
Rosso, and Gr\'egory Schehr
for organizing the summer school ``Theory of Large Deviations and Applications''
in Les Houches 2024 and for inviting me to give a series of lectures which
was the basis of the current texte.
        The author is thankful to Peter Werner for critically
        reading the manuscript. He furthermore is grateful
        to Winfried G. Schneewei\ss{} for
        introducing him to the method of biased simulations which
        happened during the
        author's Diploma thesis in Computer Science at the Fernuniversit\"at
        Hagen (Germany) in 1992. He also
        thanks Peter Grassberger for
        introducing him to the field of the statistics of biological
        sequence alignment during a workshop at Santa Barbara (USA) in 2001,
        which motivated the first physics large-deviation project
        \cite{align2002} of the author. 
        The author finally is much obliged to all his collaborators in
        the field, which have become too many to list them all here.

\bibliography{alex_refs}


\end{document}